\documentclass[preprint2]{emulateapj}

\usepackage{natbib}  
\usepackage{amssymb}

\usepackage{comment}
\usepackage{graphicx}

\bibpunct{(}{)}{;}{a}{}{,} 

\shorttitle{\Large Quark-Novae in Low-Mass X-ray Binaries}
\shortauthors{Ouyed et al.}

\begin{document}

\title{Quark-Novae in Low-Mass X-ray Binaries with massive neutron stars:\\
 a universal model for short-hard Gamma-Ray Bursts}

\author{Rachid Ouyed$^{1}$, Jan Staff$^{2}$, and Prashanth Jaikumar$^{3,4}$}

$^1$\affil{Department of Physics and Astronomy, University of Calgary, 
2500 University Drive NW, Calgary, Alberta, T2N 1N4 Canada}
$^{2}$\affil{Department of Physics and Astronomy, Louisiana State University,
202 Nicholson Hall, Tower Dr., Baton Rouge, LA 70803-4001, USA}
$^3$\affil{Department of Physics and Astronomy, California State University  Long Beach,  1250 Bellflower Blvd., Long Beach CA 90840}
$^4$\affil{Institute of Mathematical Sciences, CIT Campus, Taramani, Chennai 600113, India}

\email{rouyed@ucalgary.ca}

\begin{abstract}  We   show  that  several   features  reminiscent  of
short-hard  Gamma-ray Bursts (GRBs) arise  naturally when  Quark-Novae occur in  low-mass
X-ray  binaries born with  massive neutron stars ($\ge 1.6M_{\odot}$) and
 harboring  a circumbinary disk.  
 Near the end of the first accretion phase, conditions are  just right for  the 
 explosive conversion of the neutron star to
a quark star (Quark-Nova). In our model, the subsequent interaction of
material from  the neutron star's ejected crust  with the circumbinary
disk explains  the duration, variability and  near-universal nature of
the prompt emission in short-hard GRBs. We also describe a statistical
approach  to  ejecta  break-up  and  collision to obtain the  photon
spectrum  in our  model, which turns out remarkably similar to the empirical
Band function \citep{band93}. We apply the model to the fluence and spectrum of GRB 000727, 
 GRB 000218, and GRB980706A  obtaining excellent fits. Extended  emission (spectrum 
and duration)  is explained  by shock-heating  and ablation  of  the 
white dwarf  by  the  highly  energetic  ejecta.  Depending  on  the  
orbital separation   when  the  Quark-Nova occurs,  we isolate  
interesting regimes within  our model when  both  prompt and extended emission can  
occur. We find that the spectrum  can carry signatures typical of Type  
Ib/c SNe although these should appear less luminous than normal type Ib/c SNe. Late X-ray 
activity is due to accretion onto  the quark  star as well  as its  
spin-down luminosity. 
Afterglow activity arise  from the expanding shell 
of  material from the shock-heated  expanding circumbinary disk.   
We find a correlation between the duration  and spectrum of short-hard 
GRBs as well as modest hard-to-soft time evolution of the peak energy.
\end{abstract}
\keywords{Stars: evolution, stars: binary, stars: neutron, supernovae: general, gamma-ray burst: general}

\section{Introduction}

Intensive  sampling of  gamma-ray  burst (GRB)  light  curves by  {\it
Swift}  has revealed  that at  least  in some  short GRBs  (henceforth
shGRBs),  prompt emission  is  followed by  softer, extended  emission
lasting tens  to hundreds of  seconds. For example, the  lightcurve of
GRB 050709  \citep{villasenor05} has  a short-hard  pulse $T_{90}
\sim  0.2$s and  a  long-soft  pulse $T_{90}  \sim  130$s; GRB  050724
\citep{barthelmy05} has prominent emission lasting for $\sim 3$ s
followed by  a long,  soft, less prominent  emission peaking  at $\sim
100$  s  after  the  trigger,  while XRT  observations  reveal  strong
flare-like activities  within the first  hundreds of seconds.   In GRB
060614,  the  lightcurve  has  a  short-hard episode  followed  by  an
extended soft  emission component  with strong spectral  evolution. On
average, the principal properties  of the intense prompt component ($<
2$  s) of short  bursts with  and without  extended emission  (EE) are
indistinguishable,  suggesting  that  short bursts  are  ``universal''
\citep{norris08}.   While  the origin  of  such  extended
emission (seen roughly  in a quarter of {\it Swift}  shGRBs) is debated,
it can provide some clues to  the identity of the elusive mechanism of
shGRBs.

The association of some shGRBs  with early type galaxies that have low
star formation rates suggests that short bursts arise from a different
progenitor mechanism than long bursts and that neutron star-black hole
(NS-BH)  or double  neutron  star  (NS-NS) mergers  might  be at  play
\citep{bloom06}.  But how  does  one obtain  extended
emission in  this case? Extended  accretion episodes in  NS-BH mergers
have  been proposed  \citep{barthelmy05},  while \citet{rosswog06}
suggest that  some debris may  be launched during the  merger process,
which   would   fall   back    later   to   power   flares   at   late
times.  Alternatively,  disk  fragmentation \citep{perna06} or
magnetic field  barrier near  the accretor \citep{proga06} may
induce intermittent  accretion that power the flares.   Other types of
mergers such  as a  white dwarf-neutron star  (WD-NS) merger  \citep{king07}
have  been  advanced  to  interpret  shGRBs.  \citet{dai06}
argue that the  final product of  a NS-NS merger may  be a
heavy, differentially-rotating NS, whose post-merger magnetic activity
would give rise  to flares following the merger  events. 
 The `off axis collapsar' models \citep{lazzati10}  and the millisecond magnetar model \citep{metzger08}  offer explanations for why the extended emission from short GRBs may resemble the prompt emission from long GRBs.  However,
it is unclear how this model can explain the apparent different galactic
environments in which short and long GRBs are found.
In both types of bursts, 
the photon spectrum  is well fitted by  the Band function, hinting at
a kind of universality in the underlying physical model.

Here, we will explore an alternative model based on a Quark-nova 
explosion occurring in a low mass X-ray binary (LMXB). Such an
explosion can happen if the neutron star, in its spin-down evolution, reaches
the quark deconfinement density and subsequently undergoes a
phase transition to the more stable strange quark matter phase \citep{itoh70,bodmer71,witten84}, resulting in a conversion front that
propagates towards the surface in the detonative regime. This hypothesis is motivated by
recent work on numerical simulations of the conversion of neutron matter
to strange quark matter \citep{niebergal10b}. Previous works that have focused either exclusively on conversion
through strange matter diffusion \citep[eg.][]{Olinto87,HB88} or use pure hydrodynamics
\citep[eg.][]{Drago07,Cho94} do not find detonation, only deflagration. However,
in the most recent work on this matter \citep{niebergal10b}, we have analyzed the issue numerically including 
both reaction-diffusion {\it and} hydrodynamics consistently, albeit within a 1D approximation. We also 
included for the first time, the effects of neutrino cooling, finding
numerical evidence for laminar speeds approaching 0.04$c$ (much higher than previous works;
 $c$ is the speed of light) and a 
possible wrinkling instability in higher dimensions. As argued recently by
\citet{Hor10}, wrinkles 
cause turbulence in the conversion front that serve as a platform for detonation. If this is borne out by more 
sophisticated simulations, we will have found potentially a new engine for GRBs that 
is capable of explaining several features of the GRB light curve, as we show in this work. This exciting possibility, 
combined with the problems faced by more conventional scenarios described in the previous paragraph, should 
make our model worth exploring as a possible option.

We note that \citet{niebergal10b} is already an important step forward in resolving some of the issues with uncertainties in the 
dynamics of the Quark-Nova. The equation of state of quark matter, and hence the putative quark deconfinement
density is also unconstrained, but recently \citet{Ozel09} have argued that accurate measurement of mass and radius for a minimum of three neutron stars will provide strong constraints up to several times nuclear saturation density. While we must await future observations of neutron star parameters for the resolution of some of these issues on the underlying Quark-Nova dynamics to be more confident in our model assumptions, we nevertheless adopt a {\it working hypothesis} in this paper that the detonative regime is indeed the likely result of a quark-hadron phase transition inside a neutron star. 
From this point on, we will follow standard arguments to demonstrate that a Quark-Nova occurring  
in an LMXB following  the end of  the first  accretion phase
leads  to features  (prompt emission,  extended emission  and flaring)
reminiscent of those  discussed above in the context  of short GRBs as
observed by  {\it Swift}. The same engine at play inside a collapsar
can  naturally explain the  many similarities  between short  and long
duration GRBs,  thus exhibiting universality between the
two classes. In the end, we obtain many observed features of short GRBs
beginning with our single hypothesis of the detonative transition, implying that shGRBs could
well provide the astrophysical evidence for the Quark-nova, and that this mechanism
deserves detailed study.

This paper is organized as follows: In \S \ref{sec:qn} we describe the
LMXB  parameters  and  give  a   brief  account  of  the  Quark-Nova 
(henceforth QN)  and  the relativistic  ejecta  it produces.   In  
\S \ref{sec:circdisk},  we
describe  characteristics of the circumbinary disk (CD).
The interaction  between  the  ejecta from  the  QN and  the CD
 is studied in \S  \ref{sec:prompt} where we describe the energetics 
and spectral features of the prompt  emission.  Using our model, we perform
fits to observed GRBs that bring out the origin of  the empirical 
Band function. We also derive the duration
of  the prompt emission in our  model  and predict  an optical  counterpart.
Extended emission  arising from ablating the white  dwarf is addressed
in  \S  \ref{sec:extended}.   Late  X-ray  activity  ascribed  to  the
spin-down  evolution of  the  quark star  and  accretion from  trapped
debris is explained in \S  \ref{sec:xray} along with  afterglows from
the shocked CD. \S \ref{sec:application} contains our main conclusions.
  
\section{Quark-Nova in a NS-WD LMXB}
\label{sec:qn}

\citet{staff06}, using standard equations of state (EOS)
 of neutron-rich matter,  considered the  likely NS candidates
 that could reach  quark deconfinement  density in their core; the fiducial value was
taken to be $\sim 5\rho_0$,
where  $\rho_0\sim   2.7\times  10^{14}$g/cc  is   nuclear  saturation density. 
To reach $\sim 5\rho_{\rm 0}$, a NS 
 with gravitational mass $M_{\rm G}\sim 1.8M_{\odot}$ (using
  the APR EOS; \cite{apr98})  is required. 
   
Quark deconfinement in the core of the NS can happen in two ways:
(i)  right after the neutron star is formed, if the above criteria is satisfied. This is not relevant to the model presented, but could be important for superluminous supernovae (\citet{ouyed09,ouyed10});
(ii)   If the above criteria is not met, the neutron star can accrete mass from
a companion and reach the critical mass for deconfinement and subsequent
QN explosion. This is the scenario considered in our model.  

\subsection{The LMXB evolution}

We begin with the standard LMXB evolution model \citep{verbunt93}, where a system 
with a relatively long orbital period ($\sim$1 day) produces a recycled pulsar 
when the subgiant progenitor of the white dwarf overflows its Roche lobe. 
In the first accretion phase, mass and angular momentum losses shorten the 
orbital period. Once this first accretion phase ceases, the orbital period continues 
to decay due to gravitational wave radiation. Eventually, the orbit decays enough for 
the white dwarf itself to overflow its Roche lobe, commencing the second, ultrashort 
orbital period ($<1$ hr) LMXB phase. Here we are mainly concerned with the
first recycling phase.

The  orbital  separation of the binary, $a$, evolves according to Kepler's third law 
 \begin{equation}
 \label{eq:avsP}
 a = 2.28\times 10^9\ {\rm cm}\ (1+q)^{1/3}  \left(\frac{M_T}{M_{\odot}}\right)^{1/3} \left( \frac{P_{\rm orb.}}{60 \ {\rm s}}\right)^{2/3}\ ,
 \end{equation}
 where $q=M_{\rm WD}/M_{\rm NS}$, and $M_T=M_{\rm NS}+M_{\rm WD}$ 
 the total mass of the system.  
 The mass of the WD,  $M_{\rm WD}$, is constrained by the expected core mass of the Roche Lobe
  filling  progenitor in the LMXB phase, $M_{\rm WD}\sim 0.15 M_{\odot}$. We adopt
  $M_{\rm WD}\sim 0.1M_{\odot}$  representing the WD's fiducial value
  near the end of the first accretion phase.
 We will make use  of the WD mass-radius relation (\cite{savonjie1983}) 
 \begin{equation}
 \label{eq:wdmr}
 \frac{R_{\rm WD}}{R_{\odot}} = 0.028  (1+X)^{5/3}  M_{\rm WD, 0.1}^{-1/3} \ ,
 \end{equation}
 with $M_{\rm WD, 0.1}$ being the white dwarf mass in units of $0.1M_{\odot}$.
 Hereafter we take
  $X=0$ (the Hydrogen fraction in the WD), since  we assume  a pure  He  WD near the end  of the  first
accretion  phase.

At the end of the first recycling process (i.e. first accretion phase), the neutron star  will reach an equilibrium period
\citep{BH91} which is approximated by the Keplerian orbital period at the Alfv\'en radius
\citep{GL92}:

 \begin{equation}
 \label{eq:Pequi}
P_{\rm NS, eq.} \sim 2\, {\rm ms}\, B_{\rm NS, 9}^{6/7} R^{16/7}_{\rm NS, 6} 
M_{\rm NS, 1.4}^{-5/7} \dot{m}_{\rm NS, Edd}^{-3/7}\,, 
\end{equation}

where $B_{\rm NS, 9}$, $R_{\rm NS, 6}$ and $M_{\rm NS, 1.4}$ are the neutron star surface
magnetic dipole field, radius and mass in units of  $10^{9}$ G, $10^{6}$
cm and 1.4 solar mass respectively. 
 The accretion rate onto the neutron star, $\dot{m}_{\rm NS}$, is given 
  in units of the  Eddington limited accretion rate 
$\dot{M}_{\rm Edd}\sim10^{-8}\,M_{\sun}$\,yr$^{-1}$ above
which  the radiation  pressure generated  by accretion  will  stop the
accretion flow.

\begin{figure*}[t!]
\includegraphics[width=\textwidth]{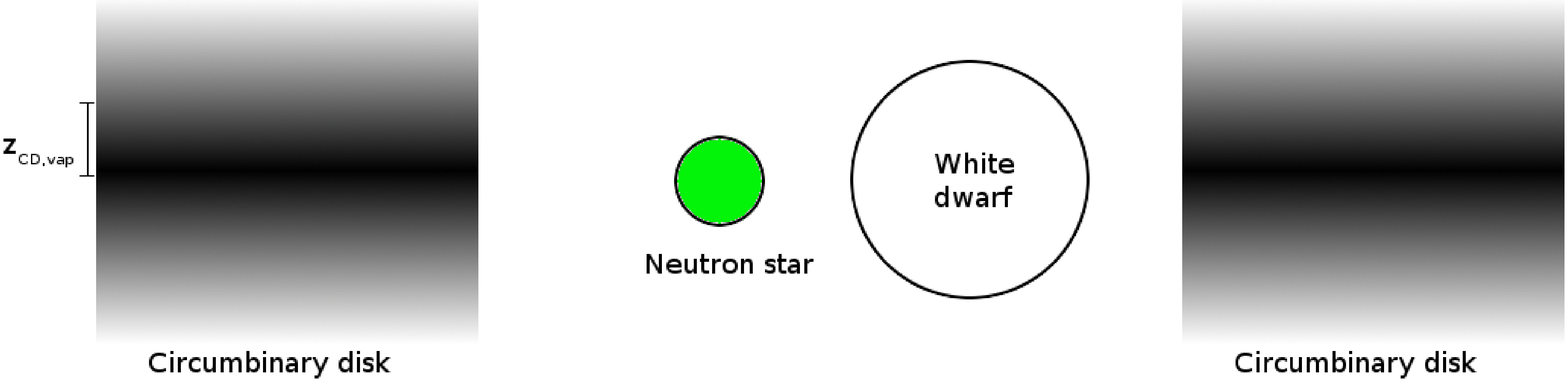}
\includegraphics[width=\textwidth]{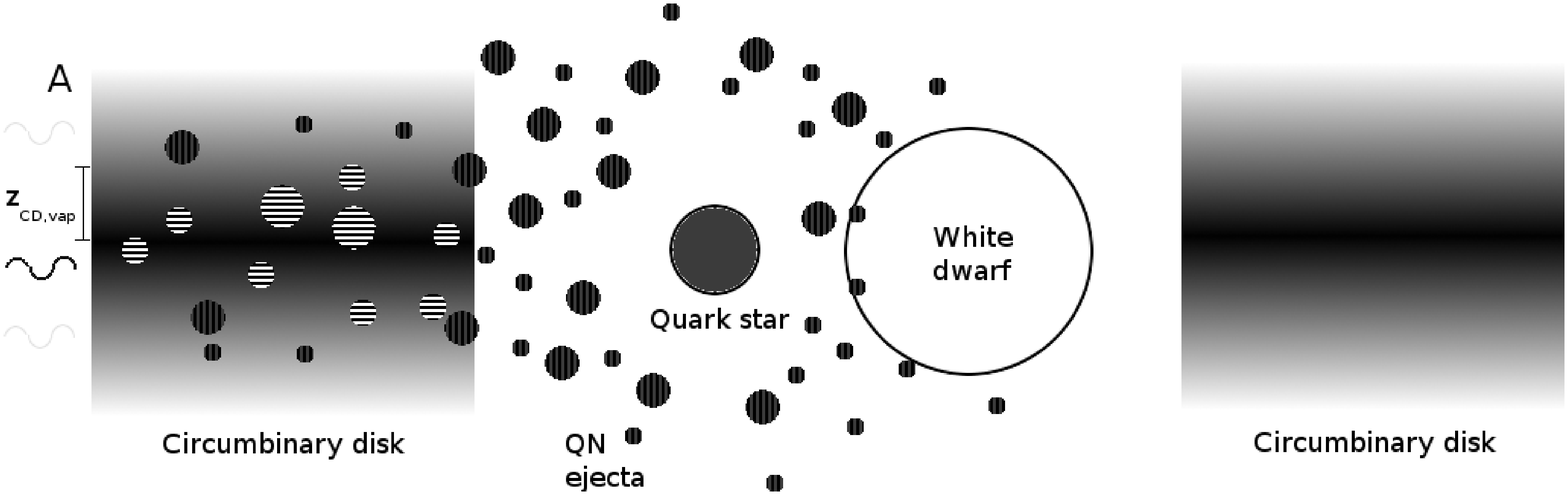}
\caption{{\bf Upper Panel: }
A LMXB system which consists of a neutron star, a low mass white 
dwarf ($M_{\rm WD} \sim 0.1 M_{\odot}$ in our model) and a
circumbinary disk  near the end of the first accretion era. 
{\bf  Lower Panel: } Near  the end of the first accretion phase, the neutron star experiences a
QN explosion ejecting the neutron star crust and leaving behind a quark
star. The QN chunks 
 are ejected during the QN explosion hitting the
circumbinary disk material at different latitudes. Below a critical height
$z_{\rm CD,vap}$ (from the midplane of the disk), the disk density is
sufficiently high that the QN chunks are shocked above
evaporation temperatures (circles with horizontal stripes). At heights greater than $z_{\rm CD,vap}$ the
density is low enough for the QN chunks to survive the shock and emit in the
optical (circles with vertical stripes). A portion of the QN chunks will collide with the white dwarf
causing it to partially or fully ablate, depending on the NS-WD separation
 and the white dwarf mass, 
when the QN explosion occurs.}
\label{schematicfig}
\end{figure*}

Neutron stars as massive as $\sim 1.8M_{\odot}$ are not easy to produce in
 LMXBs  even for  initially high mass of the donor star, unless they were already born as relatively massive neutron stars (e.g. \cite{lin2010}).
  Assuming up to $\sim 0.2M_{\odot}$ can be accreted (which
   will spin-up the NS to millisecond periods)\footnote{The  spin-up of a slowly rotating  neutron star with
moment of inertia $I_{\rm NS}$ to the
equilibrium spin period $P_{\rm NS,  eq.}=2\pi/\Omega\sim 2$ ms would 
require mass
accretion  of  at   least  $\sim  I_{\rm  NS}\Omega_{\rm  NS}/(GM_{\rm
NS}R_{\rm NS})^{1/2}\sim  0.2M_{\odot}$\citep{BH91}.}
    by the end of the first accretion phase, this puts a constraint on the
     minimum mass of the NS at birth of $\sim 1.6M_{\odot}$.
     Therefore a prolonged phase of
accretion at the Eddington rate for $> 10^7$ years is needed.    
     Thus LMXBs with NSs born massive and accreting near
      the Eddington limit are the most likely
      systems to experience a QN explosion near the end
       of the first recycling phase.    Figure \ref{schematicfig}  (upper panel)
        is a schematic representation of the LMXB system
 at the end of the first accretion phase just before the QN goes off.

\subsection{The Quark-Nova}

 In  the Quark-nova  picture (Ouyed et al. 2002),  there  is  an explosive  phase
transition  to the  more compact  {\it (u,d,s)}  quark phase,  and the
gravitational  potential  energy   is  released  partly  into  outward
propagating shock  waves from the  supersonic motion of  the ($u,d,s$)
conversion front.  The temperature of the
quark core thus formed rises quickly to 10-20 MeV since the collapse
is adiabatic rather than isothermal \citep{gentile93}.  As mentioned in the
introduction, a complete dynamical treatment  of the QN, including multi-dimensional neutrino transport and 
hydrodynamics is only  in the preliminary stages \citep{niebergal10b} 
but  results based on calculations incorporating the most physics suggest that a detonation is possible.  
The  important outcome of the QN, as we now explain,
is that  chunks of the neutron star's crustal matter, rich in iron-group
 elements, can be ejected from the surface of the 
neutron star at relativistic speeds.

\subsection{The Quark-Nova ejecta}
\label{sec:qnejecta}

Unlike Supernovae, neutrino-driven mass ejection in Quark-novae is not
feasible, as  neutrinos are trapped  inside a hot and  dense expanding
quark  core, once  it  grows  to more  than  $\sim$2~km \citep{keranen05}.
  In \citep{keranen05},  we used a typical inelastic neutrino-nucleon cross-section to show that these (electron-) neutrinos are absorbed in the neutron-rich crust of the neutron star. The corresponding diffusion timescale out of the conversion front is $\sim 0.1$ seconds.  Although the neutrinos have a similar luminosity as in supernovae (using the {\it per species} number for the latter), there is an important difference in the outcome for mass ejection. Unlike the case of a supernova, where neutrinos can drive a wind from the surface of the proto-neutron star, the neutron-rich crust is much too dense for these neutrinos to transfer kinetic energy effectively (the gravitational potential well is much deeper). Rather, they heat up the crust to temperatures of $\sim 1$ MeV. In comparison to the photon-driven  explosion \citep{vogt04} which we have invoked here, energy deposition by neutrinos is less than 10\% of the total energy budget. Therefore, it is not expected to lead to much baryon-loading of the photon fireball.  Mass ejection due to  core bounce is  also unlikely unless
the quark core is very compact (1-2) km.  A more promising alternative
is mass ejection  from an expanding thermal fireball that is a direct
consequence of dense, hot quark matter produced as a result of the
conversion from neutron matter\citep{vogt04,ouyed10}.
The  fact  that  more  than  50\%  of  the  gravitational  and
conversion energy is released as photons is unique to the QN
 \citep{ouyed05}.
It allows for  ejecta with kinetic energy easily exceeding
$10^{52}$  ergs.  Depending  on the  conversion efficiency  of photon
energy to  kinetic energy  of the ejecta  (the neutron star's outermost
layers),  up to  $10^{-2}M_{\odot}$  of neutron-rich  material can  be
ejected at  nearly relativistic speeds. Calculations  of 
mass ejection
in the QN  scenario accounting for energy transfer  to the crust, give
estimates  of ejected  mass  of $10^{-5}M_{\odot}$-$10^{-2}M_{\odot}$
 \citep{keranen05}.
The average Lorentz factor of the QN ejecta is,  
\begin{equation}
 \label{eq:gammaQN}
 \Gamma_{\rm QN}  = \frac{\eta E_{\rm QN}}{M_{\rm QN}c^{2}}
 \sim  10 \frac{E_{\rm QN,52}^{\rm KE}}{M_{\rm QN, -3.3}}\ ,
\end{equation}

where  $E_{\rm QN,  52}^{\rm KE}=  \eta_{0.1}  E_{\rm QN,  53}$ is  the
kinetic energy of the ejecta in units of $10^{52}$ erg. Here $\eta\sim
0.1$ is  the efficiency  of energy transfer  from the QN  total energy
($\sim 10^{53}$ ergs from  gravitational and phase conversion energy)
to the  ejecta's kinetic energy. The QN ejecta  mass, $M_{\rm QN,-3.3}$
is given in units of $5\times 10^{-4}M_{\odot}$ which we adopt
 as our fiducial value. 
 
 There are two main sources of uncertainty to calculating the Lorentz factor
 (the neutrino contribution, we have argued above, is not more than 10\% and
 is unlikely to change much even with improved calculations of neutrino
 transport in dense matter). The first is the thermal-to-kinetic conversion
 efficiency $\eta$ in eq.(\ref{eq:gammaQN}), which we have taken as 0.1 (e.g. \cite{frank2002}).
  Fixing this number is equivalent to fixing the amount of mass ejected.  We are at
 present working on reducing this uncertainty by numerical modeling of the
 hydrodynamics  of the conversion (from neutron to quark matter)
 front, including particle diffusion and advective
forces into account \citep[see][]{niebergal10b}. Once these simulations are
extended beyond 1D, a comparison of the thermal energy density of the
conversion front versus the pressure exerted on the surrounding matter
should provide a better constraint on $\eta$. 

 The second source of
uncertainty is the EoS and composition of the high density matter that is
ejected. Our estimate for the fiducial value of ejected mass (and hence
$\Gamma_{\rm QN}$) is based upon a fit to the EoS of the neutron star crust,
specifically the BPS equation of state for the outer crust \citep{bps71}
which is matched to a higher density equation of state for the inner crust.
The equation of state for the outer crust is reasonably robust but can be
improved by a few percent by taking into account more recent models based on
quantum molecular dynamical (QMD) simulations of the nature of the crust
post-accretion \citep[eg.][]{horowitz09}; astrophysical phenomena such as
quasi-periodic oscillations (QPOs) can also provide constraints 
\citep[eg.][]{watts07}. The higher density of the inner crust is however poorly
constrained, since it involves very neutron-rich nuclei and possibly some
``pasta" phases. We expect that upcoming experiments with rare-isotope beams
(eg. FRIB; http://www.jinaweb.org/ria/) will provide better constraints in
the high-density regime.

\citet{ol09}  studied the relativistically  expanding shell
of iron-rich ejecta emitted from  the neutron star crust and  found that it breaks
up into numerous chunks because of lateral expansion forces ($10^7$ to
$10^{11}$ chunks depending  on whether the ejecta  cools in the  solid or liquid
phase) The  typical chunk size at  birth was calculated  to be $\Delta
r_{\rm chunk, b}\sim  186\ {\rm cm} \psi_{-3}  r_{\rm b, 7}  \rho_{\rm b, 8}$
where $\psi$ is the iron  breaking strain in units of $10^{-3}$ (the iron breaking strain is $\sim
10^{-3}$  for  solid  iron and  $\sim 0.1$  for  the  liquid  phase),
  $r_{\rm b, 7}$ is the  radius in units of $10^7$
cm when breaking starts and $\rho_{\rm b, 8}$ is the ejecta  density in units 
of $10^8$ g cm$^{-3}$ at breakup \citep[see \S 3.4 and
Table 1 in][]{ol09}.  
The chunk's rest length $l_{\rm chunk}^r$ is
 much larger than its width
$\Delta r_{\rm chunk, b}\sim 200$ cm
 for  the solid phase (which we consider
  in this paper) or equal to its width for  the liquid phase \citep{ol09}.
   In the solid case, the chunks thus resemble  
 ``iron needles"  moving parallel to their long axis.
In the observer's frame the ratio of length to width will be 
contracted by $1/\Gamma_{\rm QN}$.

Even beyond the break-up radius, the chunks
remain  in  contact  with   each  other  within  the  relativistically
expanding  ejecta as a  whole. This  is because  the pieces  expand in
volume, filling up the space  between them, and causing the density of
each  piece  to  continuously  decrease,  until they  reach  the  zero
pressure iron-density  ($\rho_{\rm Fe}$),  at radius $r_{\rm sep}  \sim 2\times
10^9$ cm, at which point they stop expanding.  From Ouyed\&Leahy (2009), the  typical
 chunk size at $r_{\rm sep}$ is $\Delta r_{\rm chunk}\sim  10^5\ {\rm cm}$
  while its length is of the order of $10^6$ cm
  for the solid ejecta case. 
 As the chunks continue to expand radially outwards
 to a radius $r  > r_{\rm sep}$, we can associate with them 
a filling factor $f_{\rm chunk} =  r_{\rm sep}^{2}/r^{2}$.

We will now motivate the statistical model of ejecta break-up. 
Investigations  of the  fragment-size distribution (FSD) that results from
dynamic brittle fragmentation show  that fragmentation is initiated by
random  nucleation of  cracks that are  unstable
against  side-branch formation \citep{aastroem04}.  
The initial  cracks  and  side
branches individually form an 
exponential and  a scale invariant contribution, respectively.
Both  merge to yield  the resulting fragment  size  distribution 
which is expressed in the general form $n(s)
\propto s^{-\alpha}  f_1(s/s_1)$ with  $\alpha = (2D-1)/D$  ($D$ being
the  Euclidean dimension  of the  space), $f_1$ a  scaling  function (an
exponential) that  is independent of $s_1$ for  fragments smaller than
$s_1$ and  decays rapidly for $s>s_1$
\citep{aastroem06}. 
Here  $s_1  =
\lambda^{D}$ where  $\lambda$ is the penetration depth  of a branching
crack. Adopting this model, the emergent picture for the relativistically 
expanding shell of QN ejecta is as follows:  cracks are initially nucleated 
at random and more  or less uncorrelated  locations. From these  locations, 
cracks propagate  in  different  directions.   The  main  cracks  form  
large fragments  with  an exponential  Poisson distribution with  a
typical size determined  by $\Delta r_{\rm chunk}$.  The  size (i.e. mass)
of the  Poisson-process fragments  will be limited  or reduced  by the
creation  of small-size  fragments ($s_1$)  around each  crack  by the
side-branching process.  The final fragment  size distribution becomes
\citep{aastroem06}:

          \begin{eqnarray}
          \label{eq:fsd}
          n(s) &=&  (1-\beta) \left(\frac{s}{s_0}\right)^{-\frac{2D-1}{D}} \exp{\left(-2^D \frac{s}{s_0}\right)}\\\nonumber
&+& \beta \exp{\left[-\frac{( s^{1/D}+s_0^{1/D})^D}{s_0}\right]}\ ,
          \end{eqnarray}
          where $s_0= \Delta r_{\rm chunk}$ in our model.

\begin{figure}
\includegraphics[width=0.5\textwidth]{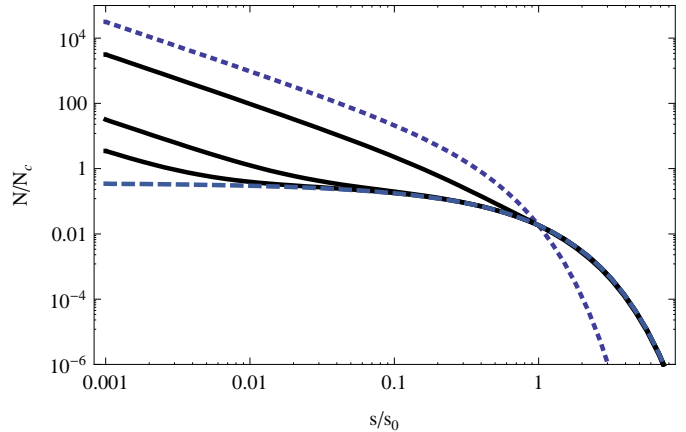}
\caption{Log-Log plot of the Quark-Nova normalized fragments size distribution
 for  $D=2$. The dotted top curve is for $\beta$=0 (chunk distribution is purely
 defined by the main cracks) while the dashed bottom curve is for 
 $\beta$=1 (chunk distribution is purely
 defined by the branching cracks). 
 The three solid curves, from top
  to bottom, are for $\beta =0.9, 0.99, 0.999$ and illustrate the contribution from the two
   size distributions as described in equation (\ref{eq:fsd}).}
   \label{fig:fsd}
\end{figure}

  Theoretical and experimental investigations \citep{sharon95,sharon96}  have shown
 that a dynamic instability controls a crack's advance when its
 velocity exceeds a critical velocity of $0.36v_{\rm R}$ where $v_{\rm R}$ is the
 Rayleigh wave speed in the material  (a Rayleigh wave is a surface
acoustic wave that travels on solids). Beyond $0.36v_{\rm R}$ 
 the mean acceleration of the crack dynamics change dramatically.
  At that point, the mean acceleration of the crack drops, the
  crack velocity starts to osciallate leading
   to the so-called Yoffe instability \citep{yoffe51} which leads to side-branching
\citep[e.g.][]{buehler06}. 
Numerical solutions of the model for $v > 0.36v_{\rm R}$ exhibit 
the occurence of these branching events where the main crack sprouts
 side branches. The spacing between these branches is a function
  of the amount of dissipation. 
 The  above equation  assumes that  crack  branches propagate
easily with large penetration depth  given by $\lambda = \Delta r_{\rm
chunk}^{1/D}$. Shown  in  Figure \ref{fig:fsd}  is the FSD
for two-dimensional fragmentation with $D=2$ (i.e.  $\alpha=1.5$)\footnote{A  special  case  is,  however,  when  a
two-dimensional  object (i.e. a  thin plate  and/or closed  shells) is
fragmented  in a  three-dimensional space.   Models revealed  that the
branching-merging  picture for  two-dimensional  fragmentation may  be
valid for  the initial stages of fragmentation,  but fragmentation may
continue  as  a  cascade  of  breakups beyond  the  limit  of  maximum
fragmentation through  branching and merging of cracks.   It was found
that at later stages additional fragments were formed with a power-law
FSD that had $\alpha\sim 1.17$ \citep{linna04}.}.  There are two
natural scales in our model, the typical fragment scale $\Delta r_{\rm chunk}$  (and thus  mass $m_{\rm  chunk}$)  and, the  scale at  which the  two
terms/contributions in the RHS of equation (\ref{eq:fsd}) are equal, which  we
call,  $\Delta r_{\rm t}$.  For $s < \Delta r_{\rm t}$ FSD is dominated
by  the side  branching. As  can be  seen in  Figure \ref{fig:fsd}   the
transition size   becomes  smaller,  i.e.  $m_{\rm t}  / m_{\rm
chunk}<<  1$   as  $\beta\rightarrow  1$.    As  we  show   later  in  \S
\ref{sec:promptspectrum},  this statistical model plays a key role in
explaining the empirically observed  Band function of GRB spectra (\cite{band93}).

\section{The circumbinary disk} 
\label{sec:circdisk}
         
Circumbinary disks  (CDs) could accompany the  formation and evolution
of  binaries. These  CDs  are  either the  remains  of fallback  disks
produced in the supernovae that  formed the compact object 
\citep[e.g.][]{wang06,cordes08},  or  material injected  into
circumbinary  orbits during  the process  of  mass loss  by the  Roche
lobe-filling  companions \citep{dubus02}.  
 The  possibility that  the CB  disk  is the
remnant of the  common envelope evolution phase of  binaries cannot be
ruled  out either.   Due to the tiny CD mass that would result, even after billions of years,  such disks 
would be difficult to detect in the optical and infrared. Nevertheless, 
the detection of excess mid-infrared flux in few LMXBs in quiescence
suggests  that dusty circumbinary material might
 be  present around some LMXBs \citep{muno2006} (see also \cite{bowler2010}). This remains 
 to be confirmed. 
 Here we consider (and assume) the  case in which the binary system is surrounded by  stable
circumbinary disk/material continuously replenished during the
process of (near-Eddington) accretion. As we have said,  building massive NSs in LMXBs would
 require near Eddington accretion rates. This, we argue, could favor 
  the formation and the sustainment of the CD.

 There are different possible configurations for the CD such as a flat
or flared disk.
 Here, we  assume the  replenished disk to be a geometrically
thin,  viscous Keplerian  circumbinary disk. Is scale height, $H_{\rm in}$,  at the inner
edge,  $R_{\rm in}$,  is $H_{\rm in}\sim  0.03 R_{\rm in}$ \citep{belle04}. 
The disks could  lie as close to the  center of mass of  the binaries as
$R_{\rm in}\simeq 1.7a$, at which point they would be tidally truncated
\citep{taam03,dubus04}.  The dependence  of the
vertically integrated  surface mass density, $\Sigma$  on radius R,
is taken  from \citep{taam01},  particularly Figure 8  in that
paper,  from which one  can approximate  
$\Sigma \sim  \Sigma_{\rm in} (R_{\rm   in}/R)$. The mass of the disk is  
then   
\begin{eqnarray}
M_{\rm
CD} &=& \int_{R_{\rm in}}^{R_{\rm out}} \Sigma(R) R dR\sim \Sigma_{\rm in}
R_{\rm    in}^2(R_{\rm    out}/R_{\rm    in}    -1)\\ \nonumber
 &\simeq&    1.5\times
10^{-6}M_{\odot} \Sigma_{\rm  in, 5} a_{10}^2  \zeta_{\rm CD, out} 
\end{eqnarray}
 where $\zeta_{\rm CD, out}=(R_{\rm out,12}/R_{\rm in,10} -1)  $ .   For a disk  that is vertically 
isothermal, at a  given radius $R$, the equation of  hydrostatic equilibrium 
combined with  the radial profile above yields \citep{belle04}:

          \begin{equation}
          \rho_{\rm CD} = \frac{\Sigma_{\rm in}}{H \sqrt{2\pi}} \left( \frac{R_{\rm in}}{R}\right)
          \exp\left(-\frac{z^2}{2 H_{\rm CD}^2}\right)\ .
          \end{equation}

Unlike  a  Shakura-Sunyaev disk,  in  CDs, the  opening
angle, $H/R$, is  expected to be roughly constant  (here $\sim 0.03$).
Recalling that $R_{\rm in}\simeq 1.7a$, the above then yields

          \begin{equation}
          \label{eq:rhoCD}
          \rho_{\rm CD} \sim \rho_{\rm in} \ 
            \left( \frac{R_{\rm in}}{R}\right)^{2}
          \exp\left(-\frac{z^2}{2 H_{\rm CD}^2}\right)\  ,
          \end{equation}

where $\rho_{\rm  in}\sim 10^{-4}\ {\rm g\  cm}^{-3}~ (\Sigma_{\rm in,
5}/a_{10})$ and  $\Sigma_{\rm in, 5}  = \Sigma_{\rm in}/10^5\  {\rm g\
cm}^2$.  The  column density  as seen at  a viewing angle  $90^{\rm o}
-i=\theta_{\rm view}  = z_{\rm  view}/R$, where $i$ is  the inclination
angle of the system is

\begin{eqnarray}
\label{eq:column}
 N_{\rm CD} &=& \frac{1}{\mu_{\rm CD}m_{\rm H}}\int_{\rm R_{\rm in}}^{\rm R}
 \rho_{\rm CD} dR \\\nonumber 
 &\sim& 6\times 10^{29}\Sigma_{\rm 5}\ \exp{\left(  -\frac{\theta_{\rm view}^2}{2\theta_{\rm c}^2}\right)} ,
\end{eqnarray}

where $m_H$ is the proton mass and $\mu_{\rm CD}\sim 1.2$ is the mean 
molecular weight for the CD material which we assume has a solar composition.

\section{Prompt emission}
\label{sec:prompt}

We first address the fate of the QN chunks when they impact the CD.
The chunks shock the CD material, and in turn, undergo a reverse shock
that heats them to a temperature found from pressure balance
   $P_{\rm chunk}= P_{\rm CD, sh.}$. The
   pressure in the shocked CD material, $P_{\rm CD, sh.}$,
    is found from the jump conditions  \citep{russo88,iwamoto89}.
    We get, at $R=R_{\rm in}$, 
\begin{equation}
\label{eq:Tchunk}
k_{\rm B} T_{\rm chunk}\sim 30 \ {\rm keV}\ \frac{\Sigma_{\rm in, 5}\Gamma_{\rm QN, 10}^2}{a_{10}} \exp\left(-\frac{z^2}{2 H_{\rm in}^2}\right) \ .
\end{equation}
  Noting that
non-degenerate iron will  vaporize if heated to $\ge  0.3$ eV \citep[][for vaporization temperature of iron at normal density]{crc973}, the
rapid fall-off of the  CD density given by eq.(\ref{eq:rhoCD}) implies
that the  vaporization is already efficently accomplished by interaction
with the inner edge of  the disk. At the inner edge of the CD, 
we can define  a   critical   disk height, $z_{\rm CD, vap.}$ (and corresponding density
$\rho_{\rm CD, vap.}$), below which the chunks will
  start vaporizing upon impact (see Fig.~\ref{schematicfig} for a
schematic illustration of the process). This critical height is  
$z_{\rm CD, vap.}\sim 4.8 H_{\rm in}$  with a corresponding CD density
  $\rho_{\rm CD, vap.}\sim 5.8\times 10^{-7}$ g cm$^{-3}$.   It
follows that the solid angle extended by the critical surface defined by 
$\rho > \rho_{\rm CD, vap.}$ is then
\begin{equation}
\Omega_{\rm c} = \frac{2\pi R_{\rm in} 2 z_{\rm CD, vap.}}{4\pi R_{\rm in}^2} \sim 4.8 \frac{H_{\rm in}}{R_{\rm in}}\sim 0.15 \  .
\end{equation}
The amount of QN ejecta  material contained within $\Omega_{\rm c}$ is $M_{\rm
QN,  c} = \Omega_{\rm  c} M_{\rm  QN}\simeq 7.5\times  10^{-5} M_{\odot}
M_{\rm  QN, -3.3}$.

 The  time it takes  a given chunk  to be completely  eroded is
\begin{equation}
t_{\rm vap.} = \frac{l_{\rm chunk}^{\rm r}}{2 v_{\rm s} \Gamma_{\rm QN}^2} \sim 4.3\times 10^{-6}\ {\rm s}\ \frac{l_{\rm chunk, 5}^{r} a_{10}^{1/2}}{\Sigma_{\rm in, 5}^{1/2}\Gamma_{\rm QN, 10}^{3}}\ ,
\end{equation}
where  $v_{\rm s} = \sqrt{(k_{\rm B}T_{\rm c}/ 2 m_{\rm H})}$ is the speed at which 
   the heat wave crosses the  chunk.
    The chunk's Thompson  optical depth, $\sim 1/(n_{\rm chunk}\sigma_{\rm T})$,  is 
  much less than the  typical chunk size, $\Delta r_{\rm chunk}\sim  10^{5}$ cm (the
  chunk's density is about $\sim 10^3$ g cm$^{-3}$ by the time they
  reach the CD; see \S 5).
  During their extremely brief life as solid entities in the CD, most of the chunks
    will emit as blackbodies at a rate $L_{\rm chunk} = A_{\rm chunk}\sigma T_{\rm chunk}^4$    where $A_{\rm chunk}\sim \pi \Delta r_{\rm chunk}^2$. 
     For a typical chunk, 
  $L_{\rm chunk}  \sim  10^{40}\ {\rm erg\ s}^{-1}\   T_{\rm c, 30}^4$ where the chunk's temperature  is given in units of 30 keV. 
 The total  luminosity in blackbody emission can be estimated by recalling 
  that the total area extended by the chunks is $\sim 4\pi a^2 $ so that 
 $L_{\rm BB}\sim \Omega_{\rm c} 4\pi a^2 \sigma T_{\rm c}^4
 \sim  10^{49}\ {\rm erg\ s}^{-1} T_{\rm c, 30}^4$. The corresponding 
   isotropic  blackbody contribution from all the chunks is   $L_{\rm BB, iso.} = L_{\rm BB}/\Omega_{\rm c} \sim  10^{50}$  erg s$^{-1}$. An observer would see a 
    blackbody  peaking at 
$\Gamma_{\rm QN} T_{\rm chunk}$.

 A given chunk will continue to plow CD material even
  after complete vaporization and erosion
   since the vaporized material will retain the directional motion and
size distribution of the incoming chunks. 
The vaporized  chunk material will eventually enter a deceleration phase 
 once it plows a mass of $\sim m_{\rm chunk}/\Gamma_{\rm QN}$  
\citep[e.g.][]{rhoads97} in CD material.  This puts
 a constraint on the minimum mass of the CD disk necessary to
  stop the QN ejecta within $\Omega_{\rm c}$. It is $M_{\rm CD, min}= M_{\rm QN, c}/\Gamma_{\rm QN}\sim 7.5\times 10^{-6}M_{\odot} M_{\rm QN, -3.3}$. 
     In our model,   this translates  to a minimum disk size $R_{\rm out, min}= R_{\rm stop}$ given by 
    $\zeta_{\rm CD, stop}\simeq R_{\rm stop}/R_{\rm in} \sim 500\ M_{\rm QN, -3.3}/(\Sigma_{\rm in, 5} a_{10}^2\Gamma_{\rm QN, 10})$.  This critical radius is important  for the kinematics and
 dynamics of a given chunk in the CD.  Assuming  
    a CD  with $R_{\rm out} > R_{\rm stop}$  implies
     that most of the vaporized QN ejecta will lose
      its momentum and kinetic energy to the CD material before it reaches the outer edge of the disk (see \S \ref{sec:promp1-E}).

 \subsection{Spectrum: Origin of the Band Function}
 \label{sec:promptspectrum}

 For the prompt emission, we isolate two broad regimes (in addition
  to the blackbody emission from the heated chunks) in our model 
depending on whether the shock temperature of the chunks exceeds 
the nuclear dissociation temperature (binding energy) for iron or not.

 \subsubsection{Regime 1 - MeV photons, transient Fe lines}
 \label{sec:prompt1}
 
 Upon impact,  the chunks are heated to temperatures
above 0.3 eV but below 8.8 MeV, in which case they vaporize without dissociation.  
Prompt emission in our model is tied to the shocked CD material, whose
temperature $T_{\rm CD,s}$ can be estimated as follows: The chunk 
 plows through the CD while  vaporizing into a hot stream
  of gaseous particles which  retain the directional motion  of the incoming chunks.
    In this case, in the observer's frame,

\begin{equation}
\label{ploweqn}
\frac{\rho_{\rm CD}}{\rho_{\rm chunk}^{\rm obs.}}\Gamma_{\rm QN}m_{\rm chunk}c^2\sim\frac{dM_{\rm CD,plow}}{\mu_{\rm CD}m_H} k_{\rm B} T_{\rm CD,s}\,.
\end{equation}

We have $dM_{\rm CD,plow}$=$A_{\rm chunk}\rho_{\rm CD}dR_{\rm CD, obs.}$ where $A_{\rm chunk}$ is the area of the chunk and $dR_{\rm CD, obs.}$ is the  distance traveled by the vaporized chunk as measured by an observer.
 The above equation is valid as long as $dM_{\rm CD, plow} < m_{\rm chunk}/\Gamma_{\rm QN}$.
The chunk's rest mass is  $m_{\rm chunk} = \rho_{\rm chunk}^{\rm r} A_{\rm chunk}  l_{\rm chunk}^{\rm r}$ with $\rho_{\rm chunk}^{\rm obs.}=\Gamma_{\rm QN} \rho_{\rm chunk}^{\rm r}$. Replacing in 
   eqn.(\ref{ploweqn})  yields

\begin{equation}
\label{eq:TCD}
 k_{\rm B} T_{\rm CD,s} \sim \mu_{CD}m_Hc^2\times \frac{l_{\rm chunk}^{\rm r}}{dR_{\rm CD, obs}}\\ .
\end{equation}
 A  limit on  the observed temperature is then,
\begin{equation}
\label{shockTcd}
k_{\rm B} T_{\rm CD, s} > 10\, {\rm keV}\, \mu_{\rm CD}\times  \frac{l_{\rm chunk, 5}^{\rm r}}{dR_{\rm CD, obs, 10}} \ ,
\end{equation}
 where $dR_{\rm CD, obs}\sim c\times  dt \approx 10^{10}$ cm to
  account for a typical GRB duration $dt_{\rm GRB}\sim 0.3$ s (see \S \ref{sec:promp1-t}). 
The temperature given above is the peak temperature after the chunk has plowed through most of the CD. Higher  peak temperatures occur in the early stages of the plowing process.

Eqn.(\ref{eq:TCD}) is illuminating - it implies a direct proportion
between the temperature of the shocked material and the linear size of the 
chunk. Since the linear size is distributed according to Eqn.(\ref{eq:fsd}), 
the temperature of the shocked CD material will also display the same 
distribution. 

It is interesting to note that several GRBs show spectra (more precisely,
$\nu F(\nu)$ curves) that appear to follow a log-normal distribution
\citep{band93}. 
According to \citet{BW95}, fragmentation of a massive 
object that proceeds by formation and branching of cracks results in
a distribution of masses (per unit logarithm in mass) that follows
the Weibull distribution~\citep{Weibull}, which looks almost identical 
to the log-normal distribution. The Weibull distribution in mass can be 
written as~\citep{BW95}

\begin{equation}
\label{eq:weibull}
 m^2n(m) = Nm_1\left(\frac{m}{m_1}\right)^{\delta+2}{\rm exp}\left[-\frac{(m/m_1)^{\delta+1}}{\delta+1}\right]
\end{equation}
where $\delta=-D/3$ is a fractal dimension and $m_1$
  a mass related to the average fragment mass in the
   distribution.  The peak of the distribution (called 
    the most probable mass; in our case  $m_{\rm chunk}$) is
     given by $m_{\rm peak}/m_1= (2+\delta)^{\frac{1}{1+\delta}}$;
     $N$ is a normalization factor related to the total number
  of fragments. 
 Note the formal similarity to the first term in eqn.(\ref{eq:fsd}), 
derived by \citet{aastroem06}, who based his expression on a more complicated fragmentation situation involving the competition between side-branching and cracking, which leads to the second term in eqn.(\ref{eq:fsd}).

\begin{figure*}[t!]
  \begin{minipage}[t]{0.5\textwidth}
    \begin{center}  
     \includegraphics[width=\textwidth]{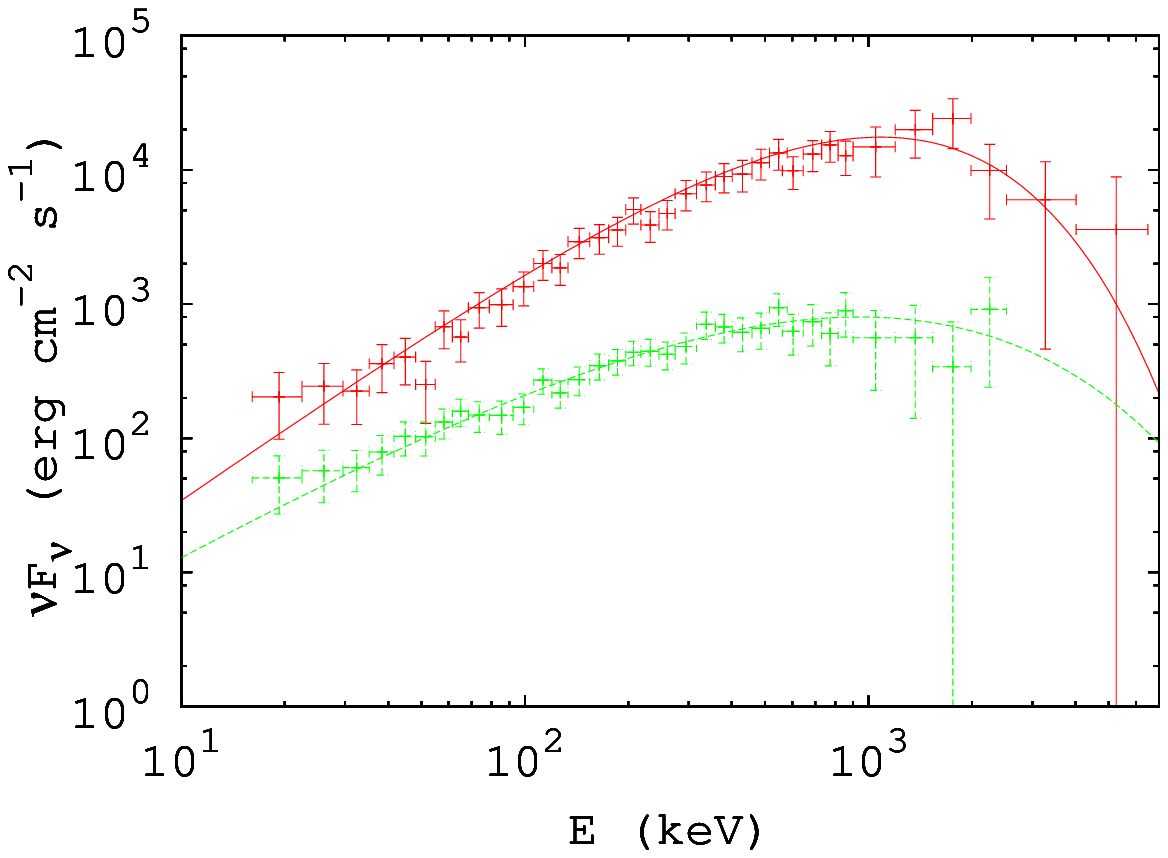}
\includegraphics[width=\textwidth]{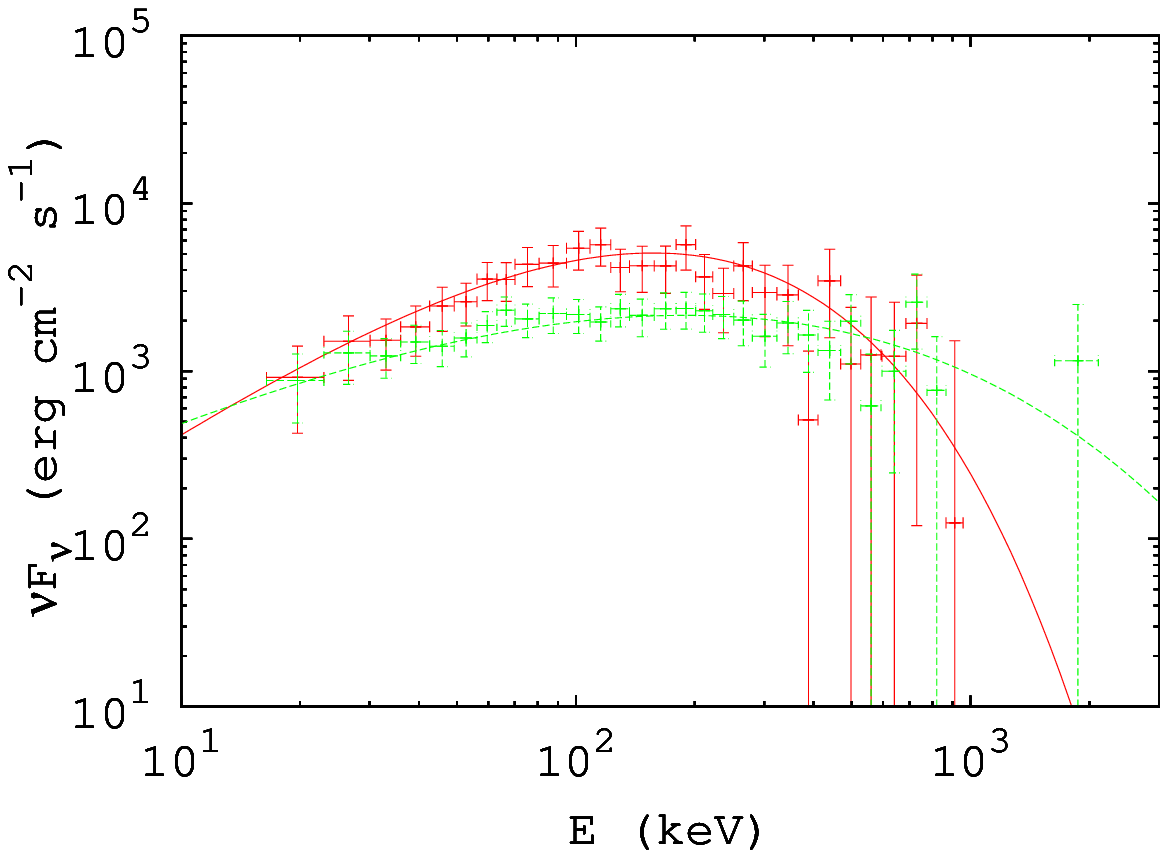}
\includegraphics[width=\textwidth]{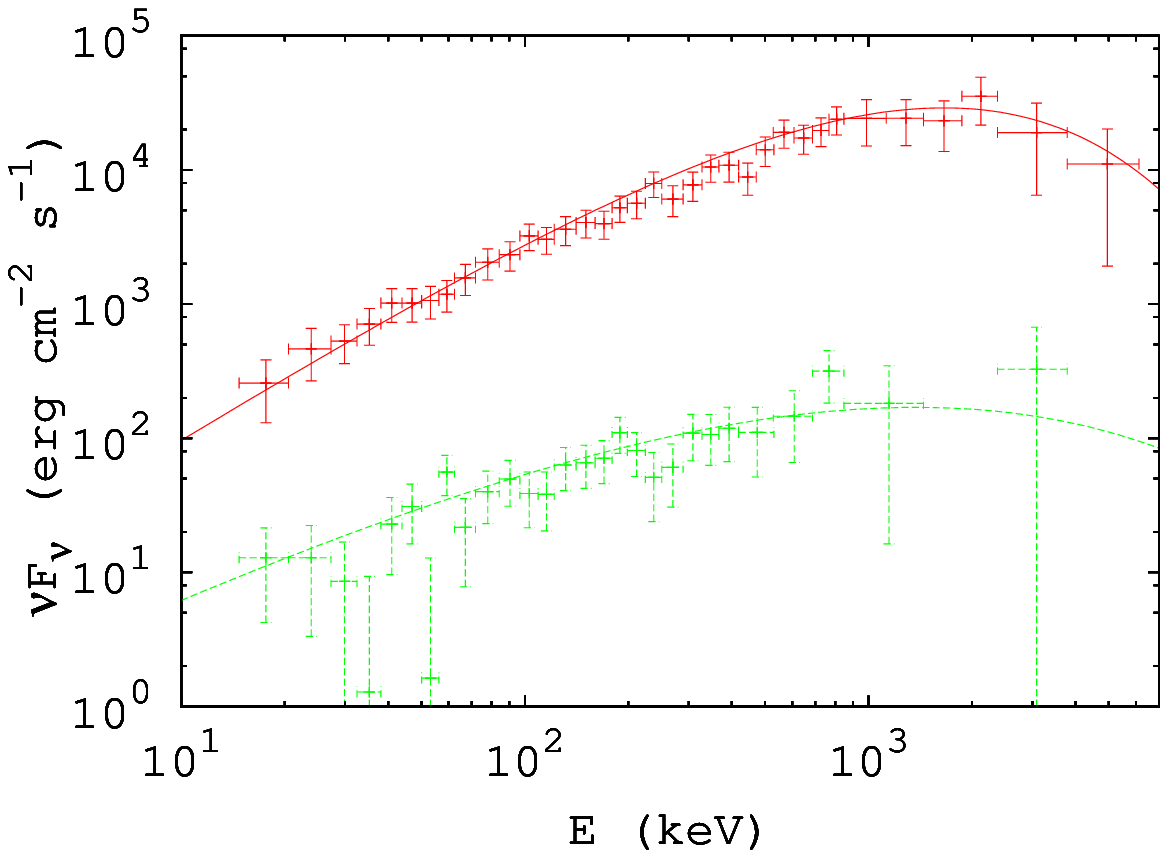}
      \caption{Model fit  (solid curves) to observed photon spectrum ($\nu F_{\nu}$) with data from
 \citet{mazets04} for GRB 000218 (top panel),  GRB 000727 (middle panel), and GRB 980706A (bottom panel). The top and bottom  data points,  in each panel,  are for different integration time (see Table 1).}
      \label{fig:nufnu}
    \end{center}
  \end{minipage}
  \hfill
  \begin{minipage}[t]{0.5\textwidth}
    \begin{center}  
     \includegraphics[width=\textwidth]{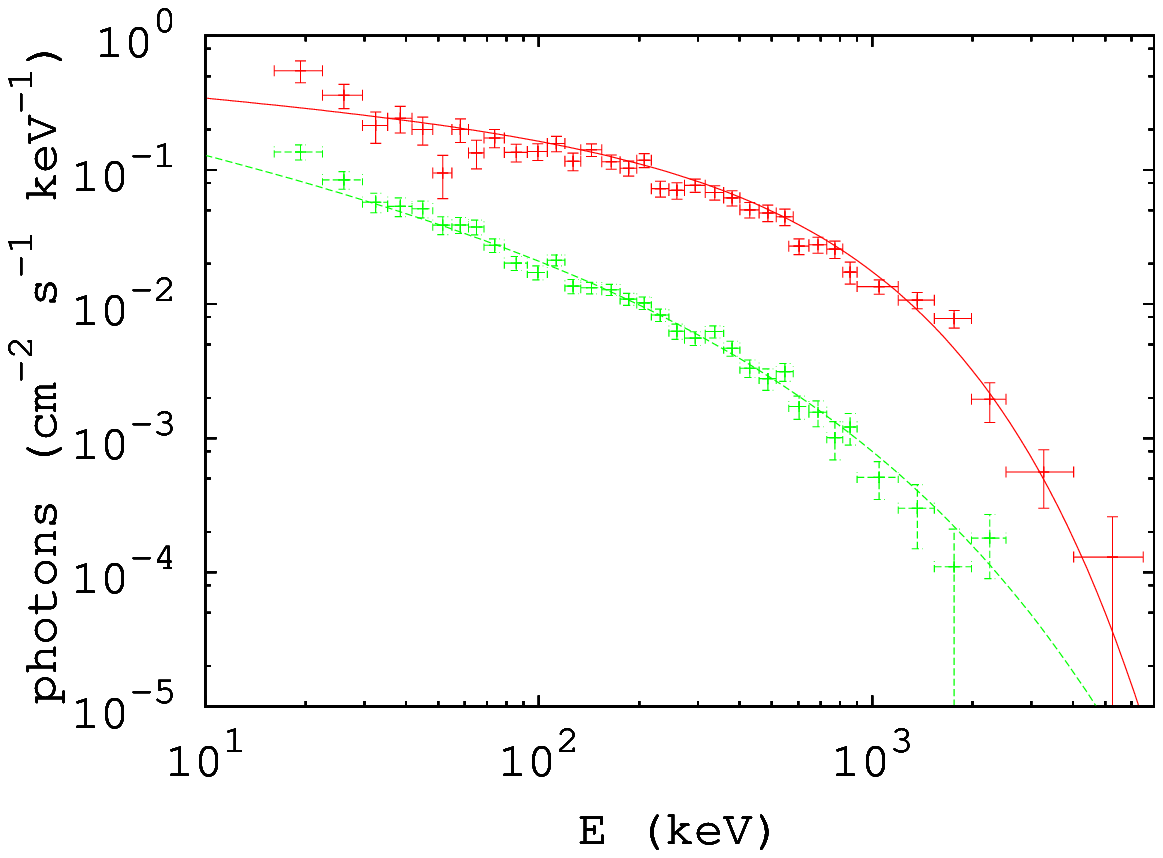}
\includegraphics[width=\textwidth]{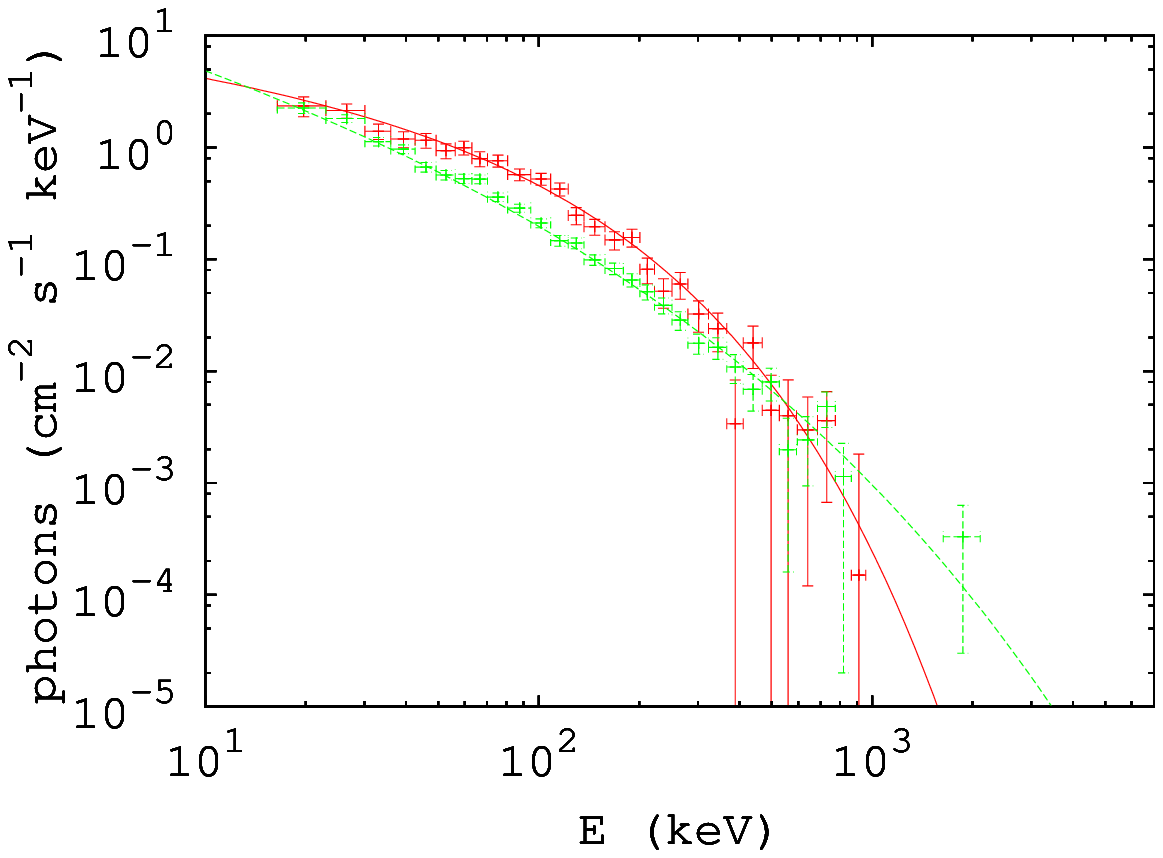}
\includegraphics[width=\textwidth]{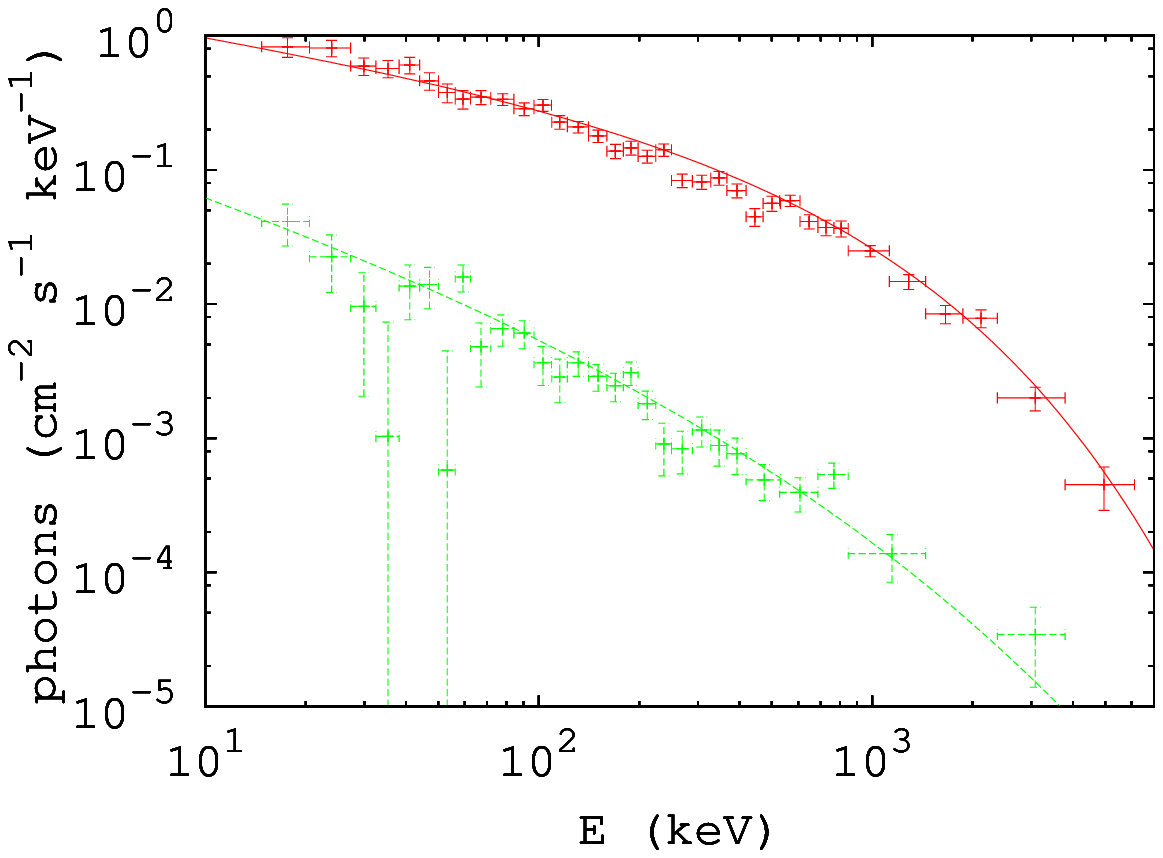}
      \caption{Model fit  (solid curves) to observed photon flux $n(E_{\gamma})$ with data from
\citet{mazets04}
  for GRB 000218 (top panel),  GRB 000727 (middle panel), and GRB 980706A (bottom panel).
  The top and bottom  data points,  in each panel,  are for different integration time (see Table 1).}
      \label{fig:fnu}
    \end{center}
  \end{minipage}
\end{figure*}

Therefore, we put forward the intriguing suggestion that the GRB
spectrum arises due to photons from the shock-heated CD material
whose temperature distribution is inherited from a fragmentation 
distribution of the QN ejecta.  We assume a one-to-one relationship
 between the temperature of the shocked CD material and the resulting
  radiation energy;   in this case, the prompt photon energy $E_{\gamma}$
   will inherit  the fragmentation distribution.
 Although a full calculation of the
non-thermal spectrum of the photons is beyond the scope of this
work, we test this idea by performing fits to some observed
GRB spectra  by assuming that the photon luminosity
   ($\nu F(\nu)= E_{\gamma}^2 n(E_{\gamma}$))
follows the Weibull distribution  given by (\ref{eq:weibull}),

\begin{equation}
\label{eq:weibullfit}
 E_{\gamma}^2 n(E_{\gamma}) = A_{\gamma}~ \left(\frac{E_{\gamma}}{E_0}\right)^{\delta+2}{\rm exp}\left[-\frac{(E_{\gamma}/E_0)^{\delta+1}}{\delta+1}\right]
\end{equation}
where $A_{\gamma}$ is a normalization factor and $E_0$ the average photon energy
in the distribution. The peak of the distribution occurs at
\begin{equation}
E_{\rm peak} = (2+\delta)^{\frac{1}{1+\delta}}~E_0\ .
\end{equation}
In our model,  the assumed one-to-one correspondence between $T_{CD,s}$ and $E_{\gamma}$ implies that $E_{\rm peak}$ evolution in time is given by eq.(\ref{eq:TCD}). 
The fits to GRB 000218, GRB 000727, and GRB 980706A  are shown in 
 Figures \ref{fig:nufnu}\&\ref{fig:fnu}.  
  In reality the spectrum evolution will be related not 
 just to the chunk distribution but also the vaporitzation rate.
 We expect the softening of the spectrum to be caused mostly by the erosion 
 of the fast moving chunks which will be heated the most and thus
  radiate the most. The total number of chunks should  also decrease in time 
   with smaller chunks most likely disappearing (i.e. fully eroded) first.
 The case of GRB 000727 is an interesting one since its observed $E_0$
  increases in time. However note the data is for
   shorter integration times than for the other two candidates
    (see Table 1). In our model,  there 
    is a minimum accumulation time (presumably of the order of milliseconds) before the  full
     mass distribution (i.e. the Band spectrum)  builds up.  
 The excellent fits suggest that the underlying physical picture relating the spectrum to the ejecta fragment size or mass distribution is probably correct.
 In reality, the fragmentation process may
be more complicated (e.g. eq.(\ref{eq:fsd})), but the Weibull distribution appears to be an excellent approximation.

There are specific predictions for this  component within our model:
 (i) The peak photon energy (related to $l_{\rm chunk}^{\rm r}$) should
  evolve in time according to equation (\ref{eq:TCD}); 
   (ii) The prompt emission  is composed of contributions 
from  $\sim 10^7$-$10^{11}$   ``blobs" of  emitting CD material
 plowed and heated by the  relativistic vaporized chunks material.
A given pulse in our model consists of different ``blobls" of heated  CD
 material  emitting simultaneously. Thus a pulse should 
     also carry the signature of the distribution in mass of the chunks  (i.e.
      the Band function); (iii) finally, the prompt emission (regime 1) 
should have signatures of iron-group elements (Fe, Co, Ni) brought in by the chunks,
but this should be transient since the  vaporized material will cool
rapidly.

\begin{table}
\begin{center}
\caption{Model  fits to observed photon spectrum (GRB 000218, GRB 000727, and GRB 980706A) using  eq.(\ref{eq:weibullfit}). The fits are performed for observations at two different integration
 time, $t_{\rm int.}$.}
\begin{tabular}{|c|c|c|c|c|c|}\hline
  GRB  & $t_{\rm int.}$  &  $\delta$  & $E_0$ (keV)   &  $E_{\rm peak}/E_0$ & $A_{\gamma}$    \\\hline\hline
   000218   &  0-256\ {\rm ms}  & -0.2  & 520   & 2.1  & 0.164    \\
      & 256\ {\rm ms}-5.888\ {\rm s}  &  -0.5 &  420  &  2.3 & 0.027    \\\hline
   000727      & 0-128\ {\rm ms}  &  -0.33 & 73   & 2.2 & 3.2    \\
         & 256\ {\rm ms}-768\ {\rm ms}  &  -0.65 & 80   &  2.4 & 5.0    \\\hline
   980706A         & 0-256\ {\rm ms}  &  -0.39 & 760   & 2.2  & 0.2    \\
         &  256\ {\rm ms}-8.448\ {\rm s} &  -0.67 & 580   &  2.4 & 0.009    \\\hline
\end{tabular}
\end{center}
\label{fittable}
\end{table}%

\subsubsection{Regime 2 - GeV photons, no Fe lines}
 \label{sec:prompt2}
 
 Chunks  heated above 8.8 MeV will be dissociated into nucleons and lose
their identity by mixing with the CD matter. 
From Eqn.(\ref{eq:Tchunk}), this regime could
occur for systems with   $\Gamma_{\rm QN} > 100$ (i.e. with $M_{\rm QN} < 5\times 10^{-5}M_{\odot}$).

The mixed
CD and QN material will have a common temperature  and 
will have some neutron excess (since the QN ejecta
inherits its composition from the neutron star crust) with 
a mean molecular weight $\mu\sim 1$. We expect
a given chunk to instantly transfer a significant fraction of its kinetic
energy into heating $m_{\rm chunk}/\Gamma_{\rm QN}$ of
CD material.  The resulting thermal energy {\it per nucleon} is then 
\begin{equation}\label{eq:tenv}
  k_{\rm B} T_{\rm nucl.} \sim  100\ {\rm GeV} \ \Gamma_{\rm QN,10}^2   \ .
\end{equation}
 
For such energies which exceed the threshold for pion 
production, processes such as $p+p\rightarrow p+p+\pi^0, n+p\rightarrow
p+p+\pi^-, p+p\rightarrow p+n+\pi^+$ can occur.  Charged pion decay
produces electrons and positrons which can annihilate to form
photons, and neutron pions can produce photons by $\pi^0\rightarrow 2\gamma$.
The  luminosity can be estimated from

\begin{eqnarray}
L&\sim&\langle E_{\pi}\rangle \frac{\rho_{\rm chunk}\rho_{\rm CD}}{m_H^2}2\pi R_{\rm in} H_{\rm CD}l_{\rm chunk}c\sigma_{pp}\\ \nonumber
&\sim& 2\times 10^{49}\,{\rm ergs/s}\, \Sigma_{\rm in, 5} l_{\rm chunk,5}a_{10}\ 
\end{eqnarray}
where  we have assumed a typical cross-section for $pp$ collisions in the tens of GeV range of $\sigma_{pp}\sim 10^{-25}$ cm$^2$
\citep[i.e. 100 millibarns,][]
{Blattnig} and a typical pion energy $\langle E_{\pi}\rangle\sim 150$ MeV (threshold production). 

We offer specific predictions for this high-energy  component within our model: (i)
 We expect this component  also to show traces of the chunk
size distribution and partly follow a log-normal-like distribution. However, 
 chunks heated to extreme temperatures ($>> 1$ GeV) will start to move away
  from the band function; (ii) it should  have no signatures of Iron-group  elements since the ejecta
is dissociated completely to nucleons; (iii) GeV and higher energy 
photons  should be  observed only in systems with $\Gamma_{\rm QN} >  100$.

For the two regimes isolated above, 
  the CD material between photons and the observer
must be optically thin.
The radiation will  scatter most off of free  electrons in the ionized
CD  material.  
The Compton  optical depth  is $N_{\rm  CD} \sigma_{\rm
C}\sim 2.7\times \Sigma_5 \exp{\left(-\theta_{\rm view}^2/2\theta_{\rm
c}^2\right)}$ where $\sigma_{\rm C}= (3/8)\sigma_{\rm T} \ln{(x)}/x^2$
is the  Compton cross-section, $\sigma_{\rm  T}= 6.652\times 10^{-25}$
cm$^2$ the  Thompson cross-section, and  $x$ is the ratio  between the
prompt photon  energy and  the electron rest-mass  which is $  \sim 1$
GeV/$(m_{\rm   e}c^2)   >>1$.   The   shocked  matter   is   therefore
Compton-optically thin to the high-energy  photons.  
 For the low energy photon regime, the Thomson optical depth is too high and this should negatively affect the ability to see the shocked chunks or the shocked disk gas  since  the deposited energy should thermalize over the whole disk. The disk geometry and vertical scale height, in our model, is assumed to be the simplest with $H/R = constant$. However other disk configurations are possible, such as a CD with a flared surface; $H \propto R^{3/2}$. This is a great advantage since a large solid angle is available for the chunks to be able to plow at an angle into increasing smaller density and thus be able to radiate in the optically thin regime. Specifically, for such a  flared disk/atmosphere most of the chunks would be flying through the larger volume of the disk/atmosphere so most energy would be deposited in the region that is dense enough to significantly slow the chunks. Up high in the flared disk/atmosphere region the density is so low that the chunks fly through unimpeded.
In summary, a $H \propto R^{3/2}$ would make a large visible impact surface available with a a longer lived black body from the optically thick part and a short lived brighter part from flared region of the disk.

\subsection{Energetics}
\label{sec:promp1-E}

 The energy deposited by the QN ejecta (into  $\sim M_{\rm QN, c}/\Gamma_{\rm QN}$
  of CD material) gives  us an estimate of the prompt GRB emission.
     The   total thermal energy produced which we assume to be roughly equal to the total GRB energy
   is  $E_{\rm GRB} \sim \Gamma_{\rm QN}\times (M_{\rm QN,c}/\Gamma_{\rm QN})
c^2 \sim  1.5\times 10^{50}\ {\rm erg}\ M_{\rm QN, -3.3}$ where we have used $M_{\rm QN,c}=\Omega_{\rm c}M_{\rm QN}$.
   The effective isotropic energy in our model is thus $E_{\rm GRB, iso.}\sim
 E_{\rm GRB}/\Omega_{\rm c} \sim  10^{51}\ {\rm erg}\ M_{\rm QN, -3.3}$.
 Observed short GRBs also have $E_{\rm GRB, iso}$ up to $10^{51} {\rm
erg}$ \citep{nakar07}.
 
 In our model, the disk only subtends an angle $\sim 15$\% of the sky.  Thus, at most $\sim 10^{51}$ ergs of the QN
 kinetic energy $10^{52}$ erg intersects the disk to be radiated. It was shown by \citep{zhang07}  
 that  the  radiative efficiency ($E_{\gamma}/(E_{\gamma}+E_{\rm K})$) could vary
 from a few percents to more than 90\% depending on how the kinetic energy is calculated.
 These studies are based on key assumptions about (i)  the unknown shock parameters;
  (ii) the required detailed afterglow modeling with its own limitation
 and (iii)  assumes the fraction of the electron energy in the internal energy of the shock to be 10\%.
 Nevertheless let us assume an efficiency of $\sim 0.1$ for short GRBs which is also the number derived in an independent study by  \citep{berger07}. A 10\% radiative efficiency would imply an  $E_{\gamma}\sim 10^{50}$ erg  in true energy release corresponding to an isotropic energy release of $E_{\gamma, {\rm iso.}}\sim 10^{51}$ erg.  This number could be higher for disks with higher solid angle (e.g. flared CDs)  and/or for QN ejecta with higher   kinetic energy.

\subsection{Duration}
\label{sec:promp1-t}
 
 In our model,  emission duration is related to the 
 time it takes the QN ejecta to be stopped by its interaction with the CD:
    \begin{equation}
\label{eq:tgrb}
t_{\rm GRB} \simeq \frac{(R_{\rm stop}-R_{\rm in})}{2c\Gamma_{\rm QN}^2}\sim 1.4\ {\rm s}\ \frac{M_{\rm QN,-3.3}}{\Sigma_{\rm in, 5} a_{10} \Gamma_{\rm QN, 10}^3} \ .
\end{equation} 
The time-variability  can vary from  microseconds to a fraction of a second 
 depending on the number of  chunks (out of the $\sim 10^7$-$10^{11}$) that emit
  simultaneously  at a given time.

 \subsection{The optical counterpart}
\label{sec:optical}
 
The QN  chunks hitting the  CD 
 in regions where $\rho_{\rm CD} < \rho_{\rm CD, vap.}$ (i.e. at heights
  $z> z_{\rm CD, vap.}$; see Fig.~\ref{schematicfig})  will not be fully
vaporized,  nor  do  they  expand significantly.   Rather,  they  pass
through the CD, effectively puncturing it.  During this interaction the
temperature of a QN chunk  is determined by shock heating,
and will not  exceed $\sim 3\ {\rm eV}\  \Gamma_{\rm QN, 10}$ as
seen by an observer; thus emission  would be in the optical band.  The
observed optical emission  is related to
shock  efficiency with  emitted  energy
$
E_{\rm  Opt.}\simeq (0.3\  {\rm
eV}\times  M_{\rm QN,  Opt.}/(\mu_{\rm e}  m_{\rm  H}))\times \Gamma_{\rm
QN}^2
\sim  10^{41}~  {\rm ergs}\ \Gamma_{\rm QN,10}^2 M_{\rm QN, -3.3}$ 
where $\mu_{\rm  e}=2$ is the mean weight per electron and $M_{\rm QN, Opt.}  =\Omega_{\rm Opt.} M_{\rm QN}$ with 
$\Omega_{\rm Opt.}=(z_{\rm CD, Opt.}- z_{\rm CD, vap.})/R_{\rm in}
\sim 0.3 H_{\rm in}/R_{\rm in}$ with the maximum height $z_{\rm CD, Opt.}$  is
 calculated from eq.(\ref{eq:Tchunk}) using the lower limit of the optical band.
Our model predicts optical  emission to appear simultaneously with the
prompt emission since the chunks  penetrate all regions of the disk at
the same  time. The optical and  prompt emission are  both composed of
short pulses  from the  impact of  the chunks striking  the CD.
    
\subsection{Afterglow activity}
\label{sec:afterglow} 

 The  innermost CD material acts  as a buffer
for the  dissociated  and pulverized  chunks.  The cross-sectional area of the chunks increases rapidly
   sweeping up  more CD mass, which leads to a run-away process. 
As they expand, they start increasing
the covering factor of the blast.  If enough chunks get dissociated
 a covering/filling factor of unity would eventually be reached.
 Energy-Momentum 
conservation implies that the mixed ejecta is ejected with 
 speed, $v_{\rm f}$,  given by 
\begin{equation}\label{eq:betagamma}
  \beta_{\rm f}\Gamma_{\rm f} = \frac{\sqrt{\Gamma_{\rm QN}^{2}-1}}{1+ \frac{M_{\rm CD}}{M_{\rm QN, c}}}\sim \frac{M_{\rm QN, c}}{M_{\rm QN, c}+M_{\rm CD}}\Gamma_{\rm QN},
\end{equation}  
where $\beta_{\rm f}=v_{\rm f}/c$.  
 Depending on $M_{\rm QN, c}/M_{\rm CD}$, 
  the mixed ejecta  resembles
a  massive shell which  expands at  mildly relativistic  speeds.
 External shocks with the circumstellar matter will slow down
further the mixed ejecta and should produce an afterglow, which could lasts for months.
  This is similar to afterglow activity in 
     the internal-external shock model of GRBs (Piran 2001)
      except that the prompt emission is caused by the chunks rather
       than by internal shocks. 
    Additional afterglow activity can also occur if the
     non-dissociated chunks 
      interact with matter beyond the outer edge of the CD.\\\\\\ 

     To summarize,  the QN naturally provides the relativistic ``bullets" required by some GRB models
\citep{hb99,um2000}. However,  the 
surrounding environment and emission region is very different in our model. The interaction between the relativistic QN chunks and
the CD (as well as the chunks' composition) is the key to the successes of our model - the origin of the Band function, as well as energetics, variability and duration which are consistent with observations.  The interaction between
 the QN chunks and the WD is another feature of our model which we explore next.

\section{Extended Emission}
\label{sec:extended} 

 Thus far,  we  have focused  on the  interaction
between the  chunks of  ejecta and the  CD as  a means to  explain the
prompt  emission.   We now  investigate  the  interaction between  the
chunks and  the WD.  As we  show below, the  WD can be ablated  by the
highly  energetic chunks,  leading  to properties  reminiscent of  the
extended emission (EE) observed in some shGRBs. 
 The density of the QN ejecta at a distance $r$ from the point of
explosion is given by  eq.(B3) in Ouyed\&Leahy (2009).  A combination of mass conservation and thermal spreading of the QN 
ejecta thickness, $\Delta r$, gives
$\rho_{\rm QN} \sim 10^3\ {\rm  g\ cm}^{-3}\  \rho_{0,14}\times \Delta r_{0, 4} / (r_{10}^{9/4} M_{\rm QN, -3.3}^{1/4})$
where the density at ejection point (the surface of the NS),  $\rho_{0}$, is given in units 
of  $10^{14}$ g cm$^{-3}$  
\citep[for an $M_{\rm QN} \sim 5\times 10^{-4} M_{\odot}$ ejecta;][]{datta95}. 
 The ejecta thickness at ejection is $\Delta r_{0}\sim 10^4$ cm or $\Delta r_{0, 4}\sim 1$
  in units of $10^4$ cm.
The chunks' density at the WD location ($r = a$) is then
 $\sim 10^3\ {\rm  g\ cm}^{-3}\ \rho_{0,14}\times \Delta r_{0, 4} / (a_{10}^{9/4} M_{\rm QN, -3.3}^{1/4})$
 which is of the order of the WD density  $\rho_{\rm WD}\sim 6.3\times 10^3\ {\rm g\ cm}^{-3}\ M_{\rm WD,0.1}^2$.
 We thus expect the QN induced shock to pass through the WD rather than  flowing around
 since the ``crushing time"  $\sim (R_{\rm WD}/c)\times (\rho_{\rm WD}/\rho_{\rm QN})^{0.5} \sim  (R_{\rm WD}/c) << $ 1 s
 \citep{klein94}.
Thus it is reasonable to assume that all of the intercepted kinetic energy
will go into heating, rather than shocking on and flowing around, the WD.

 \subsection{Energetics}   
 \label{extendedenergy}

 The total thermal energy deposited by the chunks impacting the WD is
 $\Omega_{\rm WD} E_{\rm QN}^{K}$, given by 
   
\begin{equation}
   E_{\rm th} \sim  9.3\times 10^{49}\ {\rm erg} \
    E_{\rm QN, 52}^{K}
    M_{\rm WD, 0.1}^{-2/3} a_{\rm 10}^{-2}\ , 
   \end{equation}
   
where $\Omega_{\rm WD}= R_{\rm WD}^2/(4 a^2)$ is the solid angle 
subtended by the WD. If $E_{\rm th}$ exceeds the gravitational 
binding energy of the WD 
   
\begin{equation}
   E_{\rm G, WD}=G M_{\rm WD}^2/R_{\rm WD} \sim 
   1.4\times 10^{48}\ {\rm erg}\ M_{\rm WD, 0.1}^{7/3}
   \end{equation}

then the  WD would be ablated as a result of the QN explosion.
The condition above imposes a nearness restriction on the WD as
 
\begin{equation}
  a < a_{\rm th.} = 8.2\times 10^{10}\ {\rm cm} \ \frac{{(E_{\rm QN,52}^{KE})}^{1/2}}{M_{\rm WD, 0.1}^{3/2}} \ .
\end{equation}
   
For systems with $a<  a_{\rm th.}$, one should
observe both prompt and  extended emission.

The temperature  per nucleon  of the ablated  WD will be  an important
quantity that determines the spectrum of the extended emission. It can
be estimated as

\begin{equation}
\label{eq:Textend}
k_{\rm B} T_{\rm WD} \sim 133 \ {\rm keV} \frac{\mu_{\rm WD, 4/3} E_{\rm QN, 52}^{\rm KE}}{a_{10}^2 M_{\rm WD, 0.1}^{5/3}}\ ,
\end{equation}
where $\mu_{\rm WD}$ is the mean molecular weight in units of 4/3 
(for a pure Helium WD). This implies that WD material is 
ejected at speeds 

\begin{equation}
\label{eq:Vextend}
V_{\rm th., ejec.} \sim 3\times 10^3\ {\rm km\ s}^{-1}  \frac{{E_{\rm QN, 52}^{\rm KE}}^{1/2}}{a_{10} M_{\rm WD, 0.1}^{5/6}}\ ,
\end{equation}
 
\subsection{Spectrum}

Nuclear burning timescales are much shorter than dynamical timescales.
Depending on $T_{\rm WD}$, various nuclear burning processes are expected to 
take place. As for the case of prompt emission, we can isolate the 
following interesting regimes:
  
\subsubsection{Regime 1:  $  T_{\rm WD} >  0.26\  {\rm MeV}$, Ni production}
  
From equation (\ref{eq:Textend}), this temperature regime  corresponds to

\begin{equation}
a_{10} <   a_{\rm Ni} = 0.7\ \frac{\mu_{\rm WD, 4/3}^{1/2}{E_{\rm QN, 52}^{\rm KE}}^{1/2}}{M_{\rm WD, 0.1}^{5/6}}\ ,
\end{equation}

In  this range, the  WD temperature  $ T_{\rm  WD} \ge  T_{\rm Si}\sim
3\times 10^9$  K,  the temperature  for Silicon
burning.  As a  result, $\alpha$-burning can occur all  the way up to
Nickel-56.  The sequence  of $\alpha$-burning  from Helium  through to
Nickel will  release on average 7-8  MeV per nucleon,  resulting in an
energy  release $\sim  2\times 10^{50}  \times M_{\rm  WD,  0.1}$ erg,
which  is enough  for complete  disruption  of the  WD. The  processed
nuclei are ejected at  speeds $V_{\rm nuc., ejec.}\sim 2.1\times 10^4\
{\rm km\ s}^{-1}$.  The high  speeds and nuclear processing imply that
one should observe in this regime

\begin{itemize}
\item Broad emission lines ($\sim V_{\rm nuc.,ejec.}$) early in the spectrum.
\item No Hydrogen or Helium lines, weak or no Silicon line.
\item Calcium, Nickel and Iron group lines
\end{itemize}

The signature of a QN in such  a binary system is then similar in some
respects to what is observed in Type Ic supernovae.

\subsubsection{Regime 2:  $  50\ {\rm keV} < T_{\rm WD} < 0.26\ {\rm MeV}$, Silicon production}
 
From equation (\ref{eq:Textend}), this condition corresponds to

\begin{equation}
  a_{\rm Ni}  < a_{10} <   a_{\rm Si} = 1.6 \ \frac{\mu_{\rm WD, 4}^{1/2}{E_{\rm QN, 52}^{\rm KE}}^{1/2}}{M_{\rm WD, 0.1}^{5/6}}\ ,
\end{equation}
 
This  is sufficient  for Carbon-burning  and Oxygen  burning,  but not
enough for  Silicon burning.  The observational signature  of a  QN in
such a system would be

\begin{itemize}
\item No Hydrogen or Helium  lines  
\item Oxygen, Magnesium lines
\item Strong Silicon line if Oxygen burning also occurs ($T\ge 10^9K$)
\end{itemize}

This is similar in some respects  to Type Ia SNe except that no Nickel
can be produced.  However, the QN ejecta, if  not completely destroyed
on  impact,  will  still display  weak  Fe  group  lines since  it  is
Iron-rich.  Thus,  these lines  can show up  in the  late-time spectra
making it closely resemble the characteristics of a type Ia SN.

\subsubsection{Regime 3:  $  8.6 \ {\rm keV} < T_{\rm WD} <  50\ {\rm keV} $, Oxygen  production}
 
This is sufficient for Helium ignition but not for Carbon burning.  
From equation (\ref{eq:Textend}), this regime corresponds to

\begin{equation}
 a_{\rm Si}  < a_{10} <   a_{\rm O} = 3.9 \ \frac{\mu_{\rm WD, 4}^{1/2}{E_{\rm QN, 52}^{\rm KE}}^{1/2}}{M_{\rm WD, 0.1}^{5/6}}\ ,
\end{equation}
  
 The  total nuclear  energy  released $\sim  1.8\times 10^{50}  \times
M_{\rm WD, 0.1}$  erg is once again enough  for complete disruption of
the WD, but the processed  nuclei are ejected at slower speeds $V_{\rm
nuc., ejec.}\sim 1.3\times  10^4\ {\rm km\ s}^{-1}$.  $\alpha$-burning
is the most important process  here, so that nuclei O,Ne,Mg are likely
to be synthesized. Since  Silicon burning cannot happen, elements such
as  Calcium cannot be  synthesized in  such an  explosion, but  can be
brought in  by the QN  ejecta since r-processing in  the decompressing
ejecta has  been shown to  produce abundance peaks around  Calcium and
Titanium \citep{jaikumar07}.  We expect  signatures of  such a
binary to be
  
\begin{itemize}
\item No Hydrogen, no Silicon but weak Helium lines (leftover Helium).
\item Oxygen, Magnesium lines and perhaps Calcium lines
\end{itemize}

This is similar  to the spectrum of some Type  Ic supernovae that show
weak He  lines \citep{matheson00}.   An important point  in our
model  is that Calcium  and Oxygen  originate from  different sources:
Calcium  from  slow-moving  QN  ejecta  that  do  not  get  completely
pulverized  upon impact and  Oxygen from  Helium-burning of  the white
dwarf.

\subsubsection{Regime 4:  $  T_{\rm WD} <  8.6 \ {\rm keV} $,  thermal ablation}

Nuclear burning is not possible in this system since 
Helium ignition cannot happen in this temperature regime, 
which corresponds to

\begin{equation}
 a_{10} > a_{\rm O} 
\end{equation}
 
 If  $a>a_{\rm th}$,  the extended
emission will  be weak,  but if $a<a_{\rm th}$,  thermal ablation
can still  take place, according to the  energetic arguments discussed
in \ref{extendedenergy}.  We expect signatures of a QN in 
such a binary to be
  
\begin{itemize}
\item Strong Helium lines, no Hydrogen or Silicon lines
\item $\alpha$-elements such as Calcium and Titanium
\end{itemize}

This is similar to the spectrum of Type Ib supernovae, except that
no Oxygen line should be seen. We note that the distinctions between
the regimes may not be as clear cut as presented here, rather there
should be a continuum of possibilities. A detailed calculation of
the observed spectral features is beyond the scope of this work, but
will be the subject of further studies.

 \subsection{Duration and variability}
 
 In our model, the extended emission immediately following the prompt 
emission (in the allowed regime) consists of emission of 
softer X-ray photons ($T_{\rm WD} < T_{\rm CD, obs.}$) with total energy output $\sim 10^{50}$ erg. 
Whether the WD is ablated thermally or ignited via nuclear reactions, 
the energy will be released within a dynamical time scale 
$\tau_{\rm d}\sim (G\rho_{\rm WD})^{-1/2}\sim46.5\ {\rm s}/M_{\rm WD, 0.1}$. 

The extended emission associated with short GRB tails are highly variable
and in some cases on timescales less than $\sim 1$ s 
much less than typical duration ($\sim 100$ s)  of the tail
\citep[eg.][]{norris06}. 
 Currently, it is not clear how such variability can be accounted for in our
model.
One possible future avenue is to explore  shock driven instabilities
 during the ``crushing" of the WD or mechanisms that could
  drive radial oscillations of the WD between its RL radius,
  $R_{\rm WD, RL}$, and $R_{\rm WD}$.

\section{Late X-ray activity}
\label{sec:xray}

\subsection{Nickel decay}

Systems experiencing a QN explosion while $a < a_{\rm Ni}$ 
will process most of the WD into Nickel, which will 
decay $^{56}{\rm Ni} \rightarrow ^{56}{\rm Co}$
through energetic photons with energy $\sim 0.8$ MeV \citep{pinto01}.  
The resulting energy release is 
$\sim 1.7\times 10^{48}\times M_{\rm WD, 0.1}$ erg in the X-ray.  
If released in $10^{5}$ s ($\tau_{\rm Ni}\sim 6$ days) 
this corresponds to a luminosity of $\sim 10^{43}$ erg s$^{-1}$.

\subsection{Accretion onto the Quark star}

The fate of  the WD ablated material depends on  whether it can escape
the  system. Comparing the  escape speed,  $V_{\rm esc.}  \sim 2\times
10^3\  {\rm  km\ s}^{-1}  \  M_{\rm  QS, 1.5}^{1/2}/a_{10}^{1/2}$,  to
$V_{\rm th.,  ejec.}$ and $V_{\rm  nuc.,ejec.}$ one can argue  that, in
general, part of the WD  ablated material can be trapped.  The trapped
material could then be accreted onto  the QS which could lead to X-ray
activity beyond  the EE  discussed above.  The  corresponding duration
can roughly be estimated as the free-fall time,
$\tau_{\rm ff} \sim 50\ {\rm s}\times a_{10}^{3/2}/M_{\rm QS, 1.5}^{1/2}$,
 where  the QS mass  is given  in units  of $1.5M_{\odot}$.   If enough
material is  accreted in the process,  the QS might turn  into a black
hole. The conversion of the QS to a black hole should lead to bursting
events  (X-ray spikes) occurring  shortly after  the EE.   For systems
with  $a  < a_{\rm  Ni}$  (with  signatures  of Nickel  burning),  the
free-fall  timescale  and  therefore  the duration  of  the  accretion
activity from  accretion is  brief ($ \lesssim  50$ s). They should  
also show  a late-time  ($\sim  10^5$ s)  Nickel decay  bump
lasting  for $\tau_{\rm  Ni}\sim  6$  days.  On  the  other hand,  for
$a>a_{\rm Ni}$, the duration of  the late X-ray activity could last as
long as $\sim 10^3$ seconds and no $^{56}{\rm Co}$ or $^{56}{\rm Fe}$ lines should
be present.

\subsection{Quark-star spin-down}

In the event that the QS does not turn into a black hole, it becomes a
source of late  X-ray activity.  The QN compact remnant  is a QS which
consists of  a vortex  where the interior  magnetic field  is confined
 \citep{ouyed04}.  These vortices  are expelled as  the star spins
down releasing  and dissipating its interior magnetic  field
 \citep{ouyed04,niebergal10a}.  For 
  optimum conversion of spin-down energy to radiation, the resulting X-ray luminosity
   is

\begin{eqnarray}\label{eq:qssd_lum}
L_{\rm sd}  &\sim & 3.75\times 10^{48}~{\rm erg~s}^{-1}\\\nonumber
&& \left( \frac{B_0}{10^{15}~{\rm G}}\right)^2 
    \left(\frac{2~{\rm ms}}{P_0}\right)^4 \left(1 + \frac{t}{\tau} \right)^{-5/3} \ ,
\end{eqnarray}

where $P_{0}=  P_{\rm NS, eq.}$ (see eq.(\ref{eq:Pequi})) and $B_0$ are
the  birth  spin  period  and   magnetic  field  strength  of  the  QS
respectively;  the QS magnetic field is given in units of $10^{15}$ G
as  found from studies 
of magnetic field generation in quark matter \citep[e.g.][]{iwazaki05}.
The  characteristic  spin-down  time (in  seconds)  is
\citep{niebergal06},

\begin{equation}\label{eq:charac_time}
\tau_{\rm sd} = 3.3\times 10^3~{\rm s} \left(\frac{10^{15} {\rm G}}{B_0}\right)^2 
                    \left(\frac{P_0}{2 {\rm ms}}\right)^2 
                    \left(\frac{M_{\rm QS}}{1.5M_{\odot}}\right) \left(\frac{10 {\rm km}}{R_{\rm QS}}\right)^4 \ .
\label{sdtime}
\end{equation} 

The  curve from spin-down  is flat  for up  to $\tau_{\rm  sd}$ before
decaying at a rate $t^{-5/3}$  (different from the usual $t^{-2}$ for the
neutron star  case because vortex  expulsion in the QS case  leads to  faster magnetic
field decay).  The flat segment could end abruptly in the event the QS
collapses to a  BH because of spin-down and
the subsequent increase of its core density \citep{staff06}. Finally, we note that 
the spin down energy will  be released preferably along the equator
 and should naturally be observed together with the prompt emission.

\section{Conclusions}
\label{sec:application}

In this paper we presented the idea of a Quark-Nova occurring in an LMXB 
(NS-WD binary)  during or near the end of the first accretion phase.   Our basic assumption of a detonative phase transition
occurring in the massive neutron star in an LMXB leads to many interesting consequences.
We found that several features of the short-hard GRBs
can arise naturally in our model, with the binary separation playing
an important role. Prompt emission occurs due to
the interaction between the Quark-Nova ejecta and the circumbinary disk. 
Adopting a size and speed distribution for the 
chunks can explain the range of energetic photons observed (keV-TeV)
as well as the duration  of the prompt component.
Most importantly, a statistical approach to the ejecta break-up
and collision with the  circumbinary disk could in principle lead to a light curve closely
resembling the empirical Band function. 

Extended emission in our model comes from ablation of the WD,
either thermally or via nuclear ignition. Interestingly, the spectral
features depend principally on the binary separation, and are expected
to display many similarities to type I SNe. Thus, our model offers
 a plausible unification all the Type I Supernovae.
Late X-ray activity is explained in terms of Nickel decay and
accretion of gravitationally trapped material onto the Quark star.  It
may or may  not become a black hole, and X-ray  activity can result in
either case; in  the former it will consist of  X-ray spikes while 
in the  latter, it will be  the quiescent spin-down  luminosity of the
QS. 
Afterglows  are explained in  terms of synchrotron  radiation from
the  rapidly  expanding  shocked   circumbinary disk  material.  In  all  cases,  the
energetics provide  the correct order of magnitude.

The rarity of short GRBs calls for rare situations/events. Our model
suggests Quark-Novae in LMXBs born with heavy Neutron Stars are the likely engines of the burst.
The universality of our model can be confirmed if it turns out that indeed heavy neutron stars ($>1.6M_{\odot}$) in LMXBs are  associated with or favor the formation of  circumbinary disks. The characteristics of the disk (e.g. size, density etc ...) are expected to vary slightly from one system to another  which could explain differences in burst duration, apparent energy release and spectra.

 If our model is a correct representation for short-GRBs engines, most of these engines  would reside in globular clusters (GCs) where many LMXBs are seemingly found  \citep{bogdabov06,camilo05}. It also implies that  short GRBs should be associated with early-type galaxies (no spiral arms) known to form their LMXBs in GCs.  There is some evidence that short GRBs reside in the outskirts of early-type galaxies \citep{fox07} where many GCs are located.
Finally,  the  WD ablation in our model  provides  a potential  unifying
framework for the central engines of type I Supernovae. 
In particular,  systems  observed at high latitudes (i.e. $> H/R$)
 when the QN goes off, could find some interesting applications in the context of  unusual type Ia SNe (e.g. SN2005E; \cite{perets10}) and/or subluminous type Ia SNe in general (\cite{gonzalez10}).Ó

\vskip 0.5cm

\begin{acknowledgements}
The research of R. O. is supported by an operating grant from the
National Science and Engineering Research Council of Canada (NSERC). P. J.
acknowledges start-up funds from California State University Long Beach. 
This work has been supported, in part, by grants AST-0708551, PHY-0653369,
and PHY-0326311 from the U.S. National Science Foundation and, in part, by
grant NNX07AG84G from NASA's ATP program.
J. S. and R. O. are grateful for the hospitality at California State University
Long Beach and San Diego State University during this work.
\end{acknowledgements}

\end{document}